\documentclass[12pt]{article}[amsmath,amssymb,
 aps]

\usepackage{enumitem}
\setlist{nosep}
 
\usepackage[english]{babel}
\usepackage[utf8x]{inputenc}
\usepackage{authblk}
\usepackage{graphicx}
\usepackage{dcolumn}
\usepackage{bm}
\usepackage{braket}
\usepackage{csquotes}
\usepackage{amsmath, amssymb, amsthm} 
\usepackage{apacite}
\usepackage{newtxtext}
\usepackage{newtxmath}
\usepackage{mathtools}
\usepackage{comment}
\usepackage{mathrsfs}
\usepackage{epigraph}
\usepackage{amsbsy}

\usepackage[margin=0.92in]{geometry}
\usepackage[onehalfspacing]{setspace}

\usepackage{tikz}
\usepackage{xfp} 
\usepackage[outline]{contour} 
\usetikzlibrary{decorations.markings,decorations.pathmorphing}
\usetikzlibrary{angles,quotes} 
\usetikzlibrary{arrows.meta} 
\contourlength{1.4pt}
\usetikzlibrary{snakes}

\newcommand{\calI}{\mathscr{I}} 
\tikzset{>=latex} 
\colorlet{myred}{red!80!black}
\colorlet{myblue}{blue!80!black}
\colorlet{mygreen}{green!80!black}
\colorlet{mydarkred}{red!50!black}
\colorlet{mydarkblue}{blue!50!black}
\colorlet{mylightblue}{mydarkblue!6}
\colorlet{mypurple}{blue!40!red!80!black}
\colorlet{mydarkpurple}{blue!40!red!50!black}
\colorlet{mylightpurple}{mydarkpurple!80!red!6}
\colorlet{myorange}{orange!40!yellow!95!black}
\tikzstyle{cone}=[mydarkblue,line width=0.2,top color=blue!60!black!30,
                  bottom color=blue!60!black!50!red!30,shading angle=60,fill opacity=0.9]
\tikzstyle{cone back}=[mydarkblue,line width=0.1,dash pattern=on 1pt off 1pt]
\tikzstyle{world line}=[myblue!60,line width=0.4]
\tikzstyle{world line t}=[mypurple!60,line width=0.4]
\tikzstyle{particle}=[mygreen,line width=0.5]
\tikzstyle{photon}=[-{Latex[length=4,width=3]},myorange,line width=0.4,decorate,
                    decoration={snake,amplitude=0.9,segment length=4,post length=3.8}]
\tikzstyle{singularity}=[myred,line width=0.6,decorate,
                         decoration={zigzag,amplitude=2,segment length=6.17}]

\newcommand{\If}{\mathscr{I}^+}
\newcommand{\Ip}{\mathscr{I}^-}
\newcommand{\Ifp}{\mathscr{I}^{\pm}}
\newcommand{\HS}{\mathscr{H}}
\newcommand{\BH}{\mathcal{B}}
\newcommand{\EH}{\mathcal{H}^E}

\DeclareMathOperator{\vspan}{span}

\makeatletter
\renewcommand\AB@authnote[1]{{\normalfont\textsuperscript{#1}}}
\renewcommand\AB@affilnote[1]{{\normalfont\textsuperscript{#1}}}
\makeatother

\title{The Black Hole Idealization Paradox}
\author[1]{Dominic Ryder}
\affil[1]{Department of Philosophy, Logic and Scientific Method, London School of Economics. d.ryder@lse.ac.uk}
\date{December 2024}
\begin{document}

\setlength{\abovedisplayskip}{3pt}
\setlength{\belowdisplayskip}{3pt}

\emergencystretch 3em

\maketitle
\vspace{-7ex}
\begin{center} 
Forthcoming in \textit{The British Journal for the Philosophy of Science}
\end{center}
\vspace{-2ex}

\begin{abstract}
Stephen Hawking's derivation of Hawking radiation relied on one particular spacetime model, that of a star collapsing into a black hole which then remains in existence forever. He then argued that Hawking radiation implies this model should be thrown away in favour of a different model, that of an evaporating black hole. This aspect of Hawking's argument is an example of an idealization that is pervasive in the literature on black hole thermodynamics, but which has not yet been widely discussed by philosophers. The aim of this paper is to clarify the nature of Hawking's idealization, and to show a sense in which it leads to a paradox. After identifying this idealization paradox in classic derivations of Hawking radiation, I go on to show how various research programmes in black hole thermodynamics can be viewed as possible resolutions to the paradox. I give an initial analysis of the prospects for success of these various resolutions, and show how they shed light on both the foundations of Hawking radiation and on the nature of idealizations in physics.
\end{abstract}

\newpage

\section{Introduction}\label{sec:intro}

\epigraph{He must, so to speak, throw away the ladder after he has climbed up it.}{Ludwig Wittgenstein, \textit{Tractatus Logico-Philosophicus}}

Derivations of Hawking radiation are cornerstones of modern physics. The consensus view is that Hawking radiation leads to the black hole information paradox, and huge amounts of work in physics has been dedicated to understanding and resolving it \cite{page1994black, RAJU20221}. Philosophers have also analysed various aspects of Hawking radiation, including: the black hole information paradox \cite{Belot1999, maudlin2017information, Manchak2018-MANPRA-5, wallace_2020}; black hole thermodynamics \cite{DoughertyManuscript-DOUBHT-3, Wuthrich2019, wallace2018case, WALLACE2019103, prunkl2019black}; and the universality of Hawking radiation \shortcite{Gryb2019-GRYOTU}. 

In this paper I will argue that there is another problem, distinct from those listed above, that arises because a seemingly essential idealization is used in three mainstream derivations of Hawking radiation: Hawking's original derivation \citeyear{Hawking:1975vcx}\footnote{And \textit{a fortiori} \citeA{wald1975particle}, as this is just a more mathematically rigorous version of Hawking's derivation.}, Fredenhagen and Haag's \citeyear{Fredenhagen:1989kr} ``watertight'' derivation, and algebraic approaches such as \citeA{DIMOCK1987366} and \shortciteA{dappiaggi2011rigorous}. This paper establishes the paradox for these derivations, categorises its possible resolutions, and offers an initial analysis of the success of various resolutions. The resolution of this problem, which I call the \textit{idealization paradox}, can teach us about the kinds of idealizations used in science, how global spacetime structure encodes local spacetime structure, and the nature of Hawking radiation.

The paradox arises out of an argument of \citeA{Hawking:1975vcx}, who derived the eponymous radiation in a spacetime which represents a star that collapses into a black hole which, once formed, is unchanging and exists for the rest of time. I will call this spacetime \textit{collapse-Schwarzschild}. In the same paper, Hawking also presented the first arguments that the backreaction of the radiation on the spacetime will lead to a negative energy flux into the black hole, thus causing the black hole to lose mass and evaporate. Given that a black hole evaporates, Hawking reasoned, it is not well represented by collapse-Schwarzschild. Instead, we should represent the black hole using a spacetime that models an evaporating black hole. I will this spacetime \textit{evaporation-Schwarzschild}. In other words, the use of collapse-Schwarzschild in the derivation of Hawking radiation is an idealization. Significantly, evaporation-Schwarzschild does not exhibit the same properties as collapse-Schwarzschild, and as I show in section \ref{ssec:H_Failure}, Hawking's derivation cannot be carried out in evaporation-Schwarzschild. We throw away the spacetime we were using as a ladder to Hawking radiation, collapse-Schwarzschild, in favour of evaporation-Schwarzschild, but the original derivation is not consistent with our new spacetime. 

The thesis of this paper is that, according to Hawking's and other mainstream derivations, Hawking radiation is inconsistent with black hole evaporation. It is possible to state a sketch of the paradox (the details of which I will complete in the next sections) for a derivation of Hawking radiation based upon a certain set of properties $X$:
\paragraph{\textit{The Idealization Paradox}}

\begin{enumerate}
    \item (\textbf{Hawking Radiation Derivation}) If spacetime exhibits the set of properties $X$, then Hawking radiation occurs. 
    \item (\textbf{Backreaction Arguments}) If Hawking radiation occurs, then black hole evaporation occurs. 
    \item (\textbf{Inconsistency Claim}) If black hole evaporation occurs, then spacetime does not exhibit the set of properties $X$.  
    \item (\textbf{Spacetime Postulate}) Spacetime exhibits the set of properties $X$.  
\end{enumerate}

This set of premises is inconsistent. What justifies the first premise? Of Hawking's original calculations, \citeA{unruh2014has} writes they are ``mathematically unimpeachable'', and the other derivations discussed in this paper only improve upon the degree of mathematical rigour. Thus, the secure mathematical status of the derivations in question means the first premise is hard to challenge.\footnote{Unruh also calls the calculations ``nonsense physically'' due to the \textit{trans-Planckian problem}, recently discussed by philosophers \citeA{Gryb2019-GRYOTU}. However, I ignore the trans-Planckian problem for the purpose of this paper.} What about the second premise? Using global definitions of energy one can derive a positive energy flux out toward infinity in the black hole spacetimes, as I discuss further in section \ref{ssec: HR}. Hence, assuming global energy conservation, one recovers a negative energy flux into the black hole, which is strong motivation for black hole evaporation. The third premise, Inconsistency Claim, is defended in the bulk of this paper. So what about the fourth premise? It is hard to reject Spacetime Postulate, because then we can't use the derivation of the first premise to derive Hawking radiation. So without the fourth premise, we lose our motivation for believing in Hawking radiation. Hence, according to the derivation used in the first premise, Hawking radiation is inconsistent with evaporation.

Notice that Hawking Radiation Derivation makes reference to a particular derivation.\footnote{More precisely, a particular set of properties assumed in a derivation. If two derivations of Hawking radiation assume the exact same properties, then we consider them equivalent for the purpose of this paper.} I call derivations to which the idealization paradox applies \textit{evaporation-inconsistent derivations}, and conversely those to which it does not \textit{evaporation-consistent derivations}. Thus, I argue \citeA{Hawking:1975vcx}, \citeA{Fredenhagen:1989kr} and algebraic approaches such as \citeA{DIMOCK1987366}, \shortciteA{dappiaggi2011rigorous} are evaporation-inconsistent.

Due to the possibility of evaporation-consistent derivations (and resolutions to the paradox for evaporation-inconsistent derivations), the idealization paradox does not imply we ought to be skeptical about the existence of Hawking radiation or black hole evaporation. To protect these phenomena from the paradox, one may claim that there exists a derivation of Hawking radiation that uses physically reasonable properties and is evaporation-consistent. Call this existence claim the \textit{consistency conjecture}. The `physically reasonable' qualification is necessary because a physically implausible evaporation-consistent derivation (such as a derivation in a two-dimensional spacetime) should not alleviate our concerns. I expect the consistency conjecture is true.\footnote{In the literature there are derivations which are plausible candidates for evaporation-consistency (for instance, \citeA{VISSER_2003, Parikh_2000, jacobson1991black}). I discuss these in section \ref{sec:Discussion}, but a full analysis requires another paper, which the author intends to undertake in the future of the project. See \citeA{Curiel2023} for an overview of the plethora of Hawking radiation derivations.} However, I will argue that even if the consistency conjecture is true, the idealization paradox is still paradoxical and must be resolved; the paradox identifies a mystery about how and why evaporation-inconsistent derivations were successful.\footnote{It is of course not logically necessary for a complete (consistent) theory of Hawking radiation to respond directly to the paradox. Nonetheless, our current theory of Hawking radiation admits the inconsistency so, on our current theory, the paradox requires engagement. I thank an anonymous referee for encouraging clarity on this point.}

Some philosophers, such as \citeA{Batterman2002-BATTDI, BATTERMAN2005225, Batterman2011-BATESA, Batterman2017-BATPIO} and \citeA{Morrison2012-MOREPA-4}, have argued that idealizations (construed as false descriptions) are essential for our scientific theories and models to represent and explain reality: there is ``something deeply correct about the ``unrealistic'' idealization'' \cite[p. 237]{BATTERMAN2005225}. Conversely, many have defended the view that idealizations are dispensable (for example, \citeA{Norton2012-NORAAI,butterfield2011less, Menon2013-MENCPQ} and \citeA{palacios_2019,palacios_2022}). This attitude is captured in what \citeA{jones2006ineliminable}\footnote{See also \citeA{landsman2013spontaneous, Fletcher2020-FLETPO-36}.} has called \textit{Earman's principle}: ``no effect can be counted as a genuine physical effect if it disappears when the idealizations are removed'' \cite[p. 191]{Earman2004}.\footnote{A widely discussed example in this literature is the unrealistic use of infinite limits in statistical mechanics to recover singularities in the thermodynamic theory of phase transitions. For a topical introduction to the debate and bibliography see \citeA{Shech2018-SHEIII, Shech2023-SHEIIP-4}. \shortciteA{fletcher2019infinite, Shech2018-SHEIII, Shech2023-SHEIIP-4} catalogue some of the philosophical issues that arise from the use of idealizations in physics; and see \citeA{sep-models-science, Potochnik2017-POTIAT-3} and \citeA[chapter 11]{Frigg2022} for general overviews on idealization in science.} 

Applying Earman's principle to Hawking radiation, dispensibalists will presumably demand that the collapse-Schwarzschild idealization must be removed.\footnote{Earman's principle has also been applied to Hawking radiation by \citeA{Gryb2019-GRYOTU}, in which the authors note that the response to the trans-Planckian problem which models Hawking radiation as Goldstone bosons has only been carried out in stationary spacetimes, and it is an important open question whether these models can be deidealized.} This would have the further benefit of helping to explain why new derivations of Hawking radiation continue to be produced, despite Hawking's original derivation being widely viewed as successfully establishing the phenomenon. However, as we shall soon see, deidealizing Hawking's argument is not conceptually straightforward, lending some initial plausibility to the essentialist claim. Nonetheless, a more careful look at recent research programmes in the foundations of Hawking radiation also reveals several distinct options for the dispensabilist. 

Determining how one should resolve the paradox presented here will uniquely impact our understanding of idealizations for at least two reasons. Firstly, the idealization under consideration idealizes the global structure of entire physical evolutions, which contrasts the more common-or-garden case of idealizing away certain details, for example air-resistance, and then evolving under laws from initial conditions. Secondly, usually idealized situations still obey the relevant laws; using the above example, a ball falling without air-resistance is still obeys Newton's laws. However, given one expects consistent solutions to the semi-classical Einstein equation to require black hole evaporation in the presence of Hawking radiation (due to backreaction arguments), it is not clear that the idealized model, a non-evaporation spacetime, is still consistent with the relevant laws. It remains to be established how best to treat these unique aspects of the present idealization.

My aim in this paper will be to establish the paradox for the three derivations and then categorise possible dispensabilist responses to the paradox, each of which seeks to deidealize the derivations. The plausible resolutions are associated with prominent research programmes in the foundations of Hawking radiation, including an appeal to quantum gravity, the approximation regime proposed by \citeA{Hawking:1975vcx}, and what I call ``essential structure'' derivations. I give an initial analysis of these approaches and find their prospects of success vary significantly. In particular Hawking's approximation regime fails for his own derivation, but essential structure derivations represent a very promising possible resolution. The lessons of the paradox vary across possible resolutions, but initial hints suggest insight into: the nature of Hawking radiation, how global physical properties encode local physical properties, and what sort of idealizations are used in our best physical theories.

In section \ref{sec: BH, HR and Eva} I introduce the theory of black holes and quantum field theory on black hole spacetimes that will be required. In section \ref{sec:Paradox_Hawking} I show that the idealization paradox applies to Hawking's derivation, and in sections \ref{sec: FH_paradox} and \ref{sec:Algebraic_Approach} I show that the idealization paradox bites for Fredenhagen and Haag's derivation and algebraic approaches respectively. Finally, in section \ref{sec:Discussion} I categorise and analyse resolutions to the paradox. 

\section{Primer on Quantum Field Theory in Black Hole Spacetimes}\label{sec: BH, HR and Eva}

This section reviews the background material important to the claim that the derivations I analyse are evaporation-inconsistent. I begin with black hole physics treated from the global perspective (\citeA{hawking1973large} and \citeA{Wald:1984rg}). I then introduce quantum field theory on curved spacetimes which underwrites the particle concept in Hawking radiation, before sketching the black hole evaporation heuristic. Readers familiar with quantum field theory on curved spacetime may wish to skip to section \ref{sec:Paradox_Hawking}.

\subsection{Black Hole Spacetimes and Conformal Diagrams}\label{ssec: BHs}

For my purposes, a black hole spacetime is one that is asymptotically flat at past and future null infinity ($\Ifp$) and for which there is a region of the spacetime causally isolated from the rest of the spacetime for all time ($J^-(\If) \neq M$).\footnote{Heuristically, a spacetime is asymptotically flat at $\Ifp$ \textit{iff} it is approximately Minkowski at infinity, approaches Minkowski smoothly, and is complete. For a formal definition see \citeA[chapter 11.1]{Wald:1984rg}, and see \citeA[sec. 10.3]{Landsman2021} for in depth discussion of these conditions.} The black hole region is the causally isolated region ($\BH = M - J^-(\If)$), and the event horizon bounds this region ($\EH = \dot{\BH}$). There are in fact multiple inequivalent ways to define a black hole \cite{Curiel2019-CURTMD-2}, but the one given here is standard in the formal and philosophical foundations of general relativity. Birkhoff's theorem states that any solution of Einstein's vacuum ($R_{ab} = T_{ab} = 0$) equations which is spherically symmetric\footnote{Admits the group $SO(3)$ as a group of isometries, with the group orbits spacelike two-surfaces.} in an open set $V$, is isometric in $V$ to part of the inextendible Schwarzschild solution \cite{hawking1973large}. This solution describes an uncharged, non-rotating black hole of mass $m$ with the event horizon at the Schwarzschild radius, $r= 2m$. A spacetime is said to be \textit{inextendible} if there does not exist a `larger' spacetime into which there is a proper isometric embedding. 

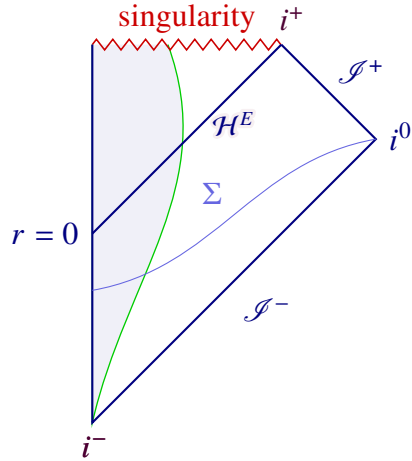
\begin{figure}
    \centering
    \begin{tikzpicture}[scale=2.5]
  \message{collapse-Schwarzschild}
  
  \coordinate (O) at (0, 0); 
  \coordinate (NE)  at ( 1, 1); 
  \coordinate (S)  at ( 0,-1); 
  \coordinate (N)  at ( 0, 1); 
  \coordinate (E)  at ( 1.5, 0.5); 
  \coordinate (X)  at ( 0.4, 1); 
  \coordinate (Y) at (0, -0.3); 

  \draw[particle, fill = mylightblue]
      (N) to (S) to[out=77,in=-70] (X);
  
  \draw[singularity] (N) -- node[above] {singularity} (NE);
  
  \draw[thick,mydarkblue] (NE) -- (E) -- (S) -- (N);
  \draw[thick,mydarkblue] (O) -- (NE);

    \path (O) -- (NE) node[mydarkblue,pos=0.75,below=4,scale=0.85]
    {\contour{mylightpurple}{$\EH$}};

  \node[above=0,left=1,mydarkblue] at (O) {$r=0$};
  \node[above=1,right=1,mydarkblue] at (E) {$i^0$};
  \node[right=1,below=1,mydarkpurple] at (S) {$i^-$};
  \node[right=4,above=1,mydarkpurple] at (NE) {$i^+$};
  \node[right=1,below=1,mydarkpurple] at (S) {$i^-$};

  \node[mydarkblue,above right=-1] at (1.25,0.75) {$\If$};
  \node[mydarkblue,below right=-1] at (0.75,-0.3) {$\Ip$};


\tikzset{mylabel/.style  args={at #1 #2  with #3}{
    postaction={decorate,
    decoration={
      markings,
      mark= at position #1
      with  \node [#2] {#3};
 } } } }
  
  \draw[world line, mylabel=at 0.5 above left with {$\Sigma$}]
      (Y) to [out = 10, in = 190] (E) ;
  
\end{tikzpicture}
    \caption{The conformal diagram for stellar collapse into a Schwarzschild black hole. The shaded region represents matter undergoing collapse.}
    \label{fig:Col_Schwarzschild}
\end{figure}

A spacetime is \textit{stationary} if it admits a global timelike Killing vector field.\footnote{A Killing vector field is a vector field whose flow is a one-parameter group of isometries $\phi_t$ ($\phi_t: M \rightarrow M$ such that $\phi_t^*g = g$).} Roughly, a stationary spacetime does not change if one follows the integral curves of the Killing vector field. Schwarzschild spacetime is stationary. However, physical black holes are formed by some astrophysical process, such as stellar collapse, so the physical spacetime will be neither stationary nor vacuum. Therefore, for a more realistic representation, we analyse a spacetime that includes spherically symmetric, non-rotating and uncharged matter that collapses into a black hole. The collapse-Schwarzschild conformal diagram is the resulting model, depicted in figure \ref{fig:Col_Schwarzschild}. This diagram will be important, so I identify some of its distinctive features. Outside the matter the spacetime is isometric to Schwarzschild by Birkhoff's theorem, and thus is stationary. Inside the matter the metric will be complicated and non-stationary. The spacetime is globally hyperbolic, meaning that it admits a Cauchy surface and thus a well-posed initial value description.\footnote{A Cauchy surface is one such that every causal curve (without an endpoint) intersects it exactly once. Therefore, heuristically, a Cauchy surface registers some information about every point in spacetime, and a globally hyperbolic spacetime is causally well-behaved. See also \citeA{Geroch1970}.} One such Cauchy surface is denoted $\Sigma$ in figure \ref{fig:Col_Schwarzschild}. Given any foliation into Cauchy surfaces, once one surface has intersected the event horizon all subsequent surfaces also will. Thus, these models describe black holes which exist forever after their formation.

The final version of a Schwarzschild black hole to consider is evaporation-Schwarzschild. However, first we need Hawking radiation.

\subsection{Quantum Fields in Black Hole Spacetimes}\label{ssec: HR}

We now turn to how particles are defined in quantum field theory, and how non-stationary spacetimes lead to particle creation. This is the core of Hawking radiation according to the mainstream view. The idea is to quantise a classical field theory by defining a Hilbert space, $\HS$, with respect to a time translation symmetry, giving a particle interpretation of the field. Because time translation symmetries are generally local in curved spacetimes, the particle interpretation is generally different in different regions, and this leads to particle creation. 

In more detail, one begins by modelling a massless complex-valued scalar field, $\Phi$, obeying the covariant wave-equation:   $g_{ab}\nabla^a\nabla^b\Phi = 0$. We can now take any of a variety of paths to define a quantum field theory, but roughly one defines a Hilbert space by selecting a subset of the solutions to the covariant wave-equation to represent physical solutions.\footnote{\label{fn:Quantization_Procedure}For example, following \citeA[p.38]{Wald:1995yp}, first define a state space for a quantum theory called a ``one-particle structure''. The covariant wave equation admits a symplectic vector space of complex-valued solutions, $(\mathscr{S}^{\mathbb{C}}, \Omega)$, where $\Omega$ is the symplectic structure on the space of solutions $\mathscr{S}^{\mathbb{C}}$. Define the Hilbert space, $\mathscr{H}$, representing physical solutions by selecting a subspace of solutions such that: (i) The ``inner product'' (scare quotes because it is not positive definite on $\mathscr{S}^\mathbb{C}$) $(y_1, y_2) = -i\Omega(\Bar{y_1}, y_2)$ is positive definite on $\mathscr{H}$, (ii) $\vspan(\mathscr{H}, \Bar{\mathscr{H}}) = \mathscr{S}^{\mathbb{C}}$, and (iii) for all $z_1 \in \mathscr{H}$ and $z_2 \in \Bar{\mathscr{H}}$, $(z_1, z_2) = 0$. Importantly, there will many choices of $\HS$ that satisfy these conditions. The Hilbert space of the full QFT will then be $\mathcal{F}_S(\HS)$, the symmetrised Fock space constructed from $\HS$.} In stationary spacetimes there is a preferred non-arbitrary way to select this subspace. Namely, we can define a global time coordinate associated with a Killing vector field that characterises time translation symmetry, and choose $\HS$ to be the space of positive frequency solutions with respect to this time coordinate (exactly as in Minkowski spacetime for an inertial time coordinate). By non-arbitrarily fixing $\HS$, we non-arbitrarily fix a particle interpretation for our QFT.\footnote{Given a positive frequency subspace with respect to a Killing vector field, time translating any state along the Killing vector field will recover a positive frequency state. Thus, the energy of the particle will always be positive when transformed by a time translation symmetry, as we desire for a particle interpretation. See \citeA[pp. 3-4]{halvorson_clifton_2002} for a brief discussion.} 
Thus, there is a preferred, global definition of a particle in stationary spacetimes. However, in general curved spacetimes there will not be a time translation symmetry which we can exploit to define positive frequency solutions. Therefore, there will not exist a non-arbitrary way to define $\HS$; so there will be no unique, global definition of a particle. This applies to collapse-Schwarzschild, which is non-stationary.

The central idea of Hawking radiation is that the failure of a spacetime to yield a global preferred particle interpretation leads to particle creation. There are local Killing vector fields at past infinity and future infinity but these differ due to non-stationarity in the bulk region of the spacetime. Therefore, given a vacuum state in the past one has particle content in the future. This is the basis of the \citeA{Hawking:1975vcx} derivation of Hawking radiation. The details are saved for section \ref{ssec:Hawking_sketch}, but in summary: we define $\HS^\pm$ on $\Ifp$ and then choose $\Phi$ such that the state is vacuum on $\Ip$; by calculating the unitary operator $U: \HS^- \rightarrow \HS^+$, we can determine the particle number for $\Phi$ on $\If$ with respect to $\HS^+$, and one finds that there is particle creation. Specifically, a thermal spectrum of particles is found at $\If$. This leads to our next topic, evaporation.\footnote{See \citeA{arageorgis2002weyling} for a challenge to the possibility of formulating unitarily implementable dynamics for quantum field theories on generic, curved spacetimes.}


Black hole evaporation cannot be directly inferred from the claim that black holes radiate, because they do not radiate like normal black bodies: no part of Hawking radiation lies in the causal future of the black hole. Instead, evaporation is thought to occur due to the backreaction of the radiation and, without a full theory of quantum gravity, this interaction between the spacetime and the radiation can't be accounted for in full rigour. One can approximate the backreaction in two ways: either by modelling the radiation as a flux going out to infinity and using a conservation law to infer a flux down over the horizon, or using the semi-classical Einstein equation, $G_{ab} = 8 \pi \langle T_{ab} \rangle$. The consensus view is that Hawking radiation implies a black hole loses mass-energy on pain of a ``drastic violation of energy conservation''\cite[pp. 282]{Fredenhagen:1989kr}.\footnote{See \citeA[sec. 7.3]{Wald:1995yp} for a treatment of the energy flux approach, and \citeA{wallace2018case} for a general overview of results in the semi-classical Einstein equation approach.} 

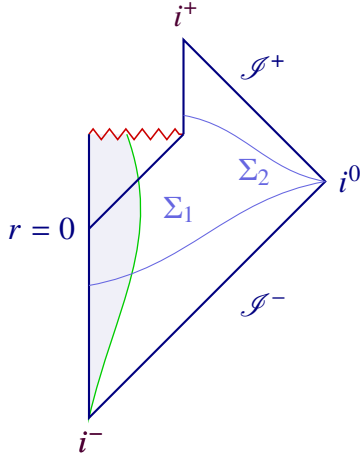
\begin{figure}
    \centering
    \begin{tikzpicture}[scale=2.5]
  \message{evaporation-Schwarzschild}
  
  \coordinate (O) at (0, 0); 
  \coordinate (NE)  at (0.5, 0.5); 
  \coordinate (NN) at (0.5, 1); 
  \coordinate (S)  at ( 0,-1); 
  \coordinate (N)  at ( 0, 0.5); 
  \coordinate (E)  at ( 1.25, 0.25); 
  \coordinate (X)  at ( 0.2, 0.5); 
  \coordinate (Y1) at (0, -0.3); 
  \coordinate (Y2) at (0.5, 0.6);
  
  \draw[particle, fill = mylightblue]
      (N) to (S) to[out=77,in=-70] (X);
  
  \draw[singularity] (N) -- node[above] {} (NE);
  
  \draw[thick,mydarkblue] (NE) -- (NN) -- (E) -- (S) -- (N);
  \draw[thick,mydarkblue] (O) -- (NE);

    \path (O) -- (NE) node[mydarkblue,pos=0.75,below=4,scale=0.85]
    {\contour{mylightpurple}{}};

  \node[above=0,left=1,mydarkblue] at (O) {$r=0$};
  \node[above=1,right=1,mydarkblue] at (E) {$i^0$};
  \node[right=1,below=1,mydarkpurple] at (S) {$i^-$};
  \node[right=1,above=1,mydarkpurple] at (NN) {$i^+$};
  \node[right=1,below=1,mydarkpurple] at (S) {$i^-$};

  \node[mydarkblue,above right=-1] at (0.75,0.75) {$\If$};
  \node[mydarkblue,below right=-1] at (0.75,-0.3) {$\Ip$};


\tikzset{mylabel/.style  args={at #1 #2  with #3}{
    postaction={decorate,
    decoration={
      markings,
      mark= at position #1
      with  \node [#2] {#3};
 } } } }
  
  \draw[world line, mylabel=at 0.5 above left with {$\Sigma_1$}]
      (Y1) to [out = 10, in = 190] (E) ;

    \draw[world line, mylabel=at 0.5 below with {$\Sigma_2$}]
      (Y2) to [out = -10, in = 170] (E) ;
  
\end{tikzpicture}
    \caption{The conformal diagram representing the evaporation of a Schwarzschild black hole formed by collapse. The mass of the black hole is shrinking over time, and after the evaporation, the spacetime is locally Minkowski. Neither $\Sigma_1$ nor $\Sigma_2$ is a Cauchy surface.}
    \label{fig:Eva_Scwarzschild}
\end{figure}

The above approaches imply the black hole will radiate away all of its mass in finite time. The semi-classical approximation is expected to break down at late times, when the radius of the black hole is of the order of the Planck length. Beyond this point there much disagreement about the description of Hawking radiation and evaporation. However, the consensus has varied very little from Hawking's original heuristic: ``there is not much it can do except disappear altogether.'' \cite[pp. 219]{Hawking:1975vcx} Thus, black holes are usually supposed to evaporate entirely, with the spacetime in the region after evaporation isometric to a region of Minkowski spacetime. The conformal diagram for this spacetime, which I call evaporation-Schwarzschild, is depicted in figure \ref{fig:Eva_Scwarzschild}.

Since this spacetime will be central to my discussion, I will highlight a few important features of it. It is very different from collapse-Schwarzschild: the metric in the region exterior to the collapsing matter is not Schwarzschild, it is not globally hyperbolic, it does not admit a timelike Killing vector field and it has a naked singularity, among other things. A consequence of Hawking radiation is that collapse-Schwarzschild is the wrong spacetime to describe the target black hole; it is an idealization. Given Hawking radiation, an uncharged, non-rotating black hole should be described by evaporation-Schwarzschild. And yet, Hawking derived the eponymous radiation in collapse-Schwarzschild, in spite of the fact that many properties of collapse-Schwarzschild that are used in Hawking's derivation do not hold in evaporation-Schwarzschild; this threatens an essential idealization. Thus we arrive at the paradox discussed in the introduction. The rest of this paper defends the claim that throwing away the ladder of collapse-Schwarzschild really leads to inconsistency.

\section{Idealization Paradox in Hawking's Derivation}\label{sec:Paradox_Hawking}

\subsection{Sketch of Hawking's Derivation}\label{ssec:Hawking_sketch}

To understand exactly what goes wrong for Hawking's derivation in evaporation-Schwarzschild, we will need a more precise account of it. I give this here, stripped of unnecessary details.

In outline, we wish to compare the modes of a quantum field in the distant past with those in the distant future. Consider collapse-Schwarzschild spacetime\footnote{Hawking also derived the radiation for charged, rotating black holes, but I focus on the simplest case.} containing a massless complex-valued scalar quantum field $\Phi$ (obtained as discussed above). Let $\{f_i\}$ be a complete basis of solutions, so that we may write: $\Phi = \sum_i \{f_i \mathbf{a}_i + \bar{f}_i \mathbf{a}_i^\dagger\}$, where $\mathbf{a}_i$ and $\mathbf{a}_i^\dagger$ are the annihilation and creation operators corresponding to the $i$th solution. We choose $\{f_i\}$ to be positive frequency solutions with respect to a time parameter defined by a timelike Killing vector field asymptotically close $\Ip$. 

We can also describe $\Phi$ as a decomposition into solutions at $\If$ and on the event horizon $\EH$. At $\If$ we can again form a Hilbert space generated by positive frequency solutions, $\{p_i\}$, with respect to a time parameter defined by a timelike Killing vector field on $\If$. The modes on $\EH$ play no role in the derivation. $\mathbf{b}_i$, $\mathbf{b}_i^\dagger$ are the annihilation and creation operators for the $p_i$ modes. Because the spacetime is globally hyperbolic, we can express $\{p_i\}$ and $\mathbf{b}_i$ as linear combinations of $\{f_i\}$ and $\{\bar{f}_i\}$ and $\mathbf{a}_i$ and $\mathbf{a}_i^\dagger$ respectively, \begin{align}\label{eq:latetimesolutiondecomposition}
   p_i = \sum_j \{\alpha_{ij}f_j  + \beta_{ij}\bar{f}_j\}, && \mathbf{b}_i = \sum_j \{\bar{\alpha}_{ij}\mathbf{a}_j - \bar{\beta}_{ij}\mathbf{a}_j^\dagger \}
\end{align}
Stipulate that the field is in the state $|0_-\rangle$ defined as the vacuum state at early times: $\mathbf{a}_i |0_-\rangle = \mathbf{0} \hspace{2mm}$ for all $i$. On $\If$, $\mathbf{b}_i^\dagger \mathbf{b}_i$ has expectation value \begin{equation}\label{eq:scri+numberopexpectationvalue}
    \langle 0_-|  \mathbf{b}_i^\dagger \mathbf{b}_i |0_- \rangle = \sum_j |\beta_{ij}|^2 
\end{equation}
which will be non-zero because we have different Killing vector fields defining our Hilbert spaces. Thus, to determine the expected number of particles in each mode, one needs to determine the coefficients $\beta_{ij}$.

For this calculation, Hawking writes the modes of $\Phi$ in terms of advanced and retarded Eddington-Finkelstein coordinates:\begin{align}\label{eq:EFcoords}
    v = t + r_*, &&  u = t - r_*, &&   r_* = r + 2m \log |\frac{r}{2m} - 1|
\end{align}

Hawking considers a mode $p_i$ on $\If$ at late retarded time $u$ of frequency $\omega$, defined with respect to retarded time, $p_\omega(u)$. He propagates this mode back along the event horizon through the non-stationary region of the collapsing star onto the $\Ip$ (see figure \ref{fig:HawkingRadiation}). The form of the mode on $\Ip$ is determined by connecting the mode to the event horizon by a null vector normal to the horizon, and parallel transporting this vector onto $\Ip$.\footnote{\label{fn:Maximal_isometry}In fact Hawking conducts the calculation on the past horizon of maximally extended Schwarzschild and argues that the conclusions would be the same on $\Ip$.} From the form of the mode on $\Ip$, one can read off the $\beta$ coefficients.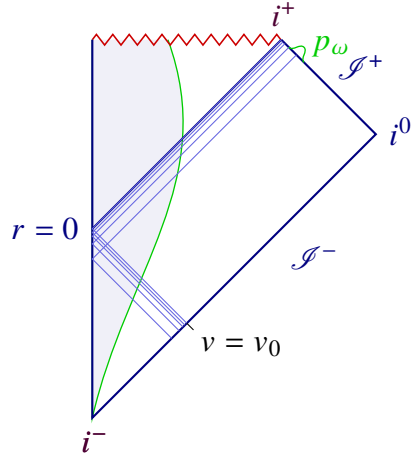
\begin{figure}
    \centering
    \begin{tikzpicture}[scale=2.5]
  \message{Hawking Derivation}

  
  \coordinate (O) at (0, 0); 
  \coordinate (NE)  at ( 1, 1); 
  \coordinate (S)  at ( 0,-1); 
  \coordinate (N)  at ( 0, 1); 
  \coordinate (E)  at ( 1.5, 0.5); 
  \coordinate (X)  at ( 0.4, 1); 
  
  \coordinate (O1) at (0, -0.02); 
  \coordinate (O2) at (0, -0.04); 
  \coordinate (O3) at (0, -0.08); 
  \coordinate (O4) at (0, -0.16); 
  \coordinate (F1) at (1.01, 0.99); 
  \coordinate (F2) at (1.02, 0.98); 
  \coordinate (F3) at (1.04, 0.96); 
  \coordinate (F4) at (1.08, 0.92); 
  \coordinate (PO) at (0.5, -0.5);
  \coordinate (P1) at (0.49, -0.51); 
  \coordinate (P2) at (0.48, -0.52); 
  \coordinate (P3) at (0.46, -0.54); 
  \coordinate (P4) at (0.42, -0.58); 

  \coordinate (N1) at (1.04, 0.96);
  \coordinate (N2) at (1.12, 0.96);
  \coordinate (N3) at (1.12, 0.88);

  \draw[particle, fill = mylightblue]
      (N) to (S) to[out=77,in=-70] (X);
  
  \draw[singularity] (N) -- node[above] {} (NE);
  
  \draw[thick,mydarkblue] (NE) -- (E) -- (S) -- (N);
  \draw[thick,mydarkblue] (O) -- (NE);

    \path (O) -- (NE) node[mydarkblue,pos=0.75,below=4,scale=0.85]
    {\contour{mylightpurple}{}};

  \node[above=0,left=1,mydarkblue] at (O) {$r=0$};
  \node[above=1,right=1,mydarkblue] at (E) {$i^0$};
  \node[right=1,below=1,mydarkpurple] at (S) {$i^-$};
  \node[right=1,above=1,mydarkpurple] at (NE) {$i^+$};
  \node[right=1,below=1,mydarkpurple] at (S) {$i^-$};

  \node[mydarkblue,above right=-1] at (1.25,0.75) {$\If$};
  \node[mydarkblue,below right=-1] at (1,-0.05) {$\Ip$};


\tikzset{mylabel/.style  args={at #1 #2  with #3}{
    postaction={decorate,
    decoration={
      markings,
      mark= at position #1
      with  \node [#2] {#3};
 } } } }

\draw[world line]
   (NE) to (O) to (PO) ;
  \draw[world line]
   (F1) to (O1) to (P1);
   \draw[world line]
   (F2) to (O2) to (P2);
   \draw[world line]
   (F3) to (O3) to (P3);
   \draw[world line]
   (F4) to (O4) to (P4);
 
  \draw[particle, mylabel=at 0.4 right with {$p_\omega$}]
  (N1) to[out=-45, in=135] (N2) to[out=-45, in=135] (N3);

    \draw[mylabel=at 0.4 below right with {$v=v_0$}]
    (PO) to ++ (-45:0.05) ;

\end{tikzpicture}
    \caption{Conformal diagram used to visualise the mapping of modes of a quantum field on $\If$ to modes on $\Ip$, which pass through the non-stationary region of collapsing matter.}
    \label{fig:HawkingRadiation}
\end{figure}
Thus we arrive at Hawking's discovery: the expected particle number at frequency $\omega$ at $\If$ is that of a black body with temperature, in geometric units, of $\frac{\kappa}{2\pi}$, where $\kappa$ is the surface gravity of the black hole. The black hole is seemingly radiating at what is now called the Hawking temperature.

Our task now is to investigate why this derivation cannot be carried out in evaporation spacetime. I begin by identifying a globally hyperbolic sub-spacetime of evaporation Schwarzschild that might plausibly admit Hawking's derivation. I then show that requisite structure used in Hawking's derivation is not present in evaporation spacetimes and so Hawking's derivation is evaporation-inconsistent.

\subsection{Hawking's Derivation Fails in Evaporation-Schwarzschild}\label{ssec:H_Failure}

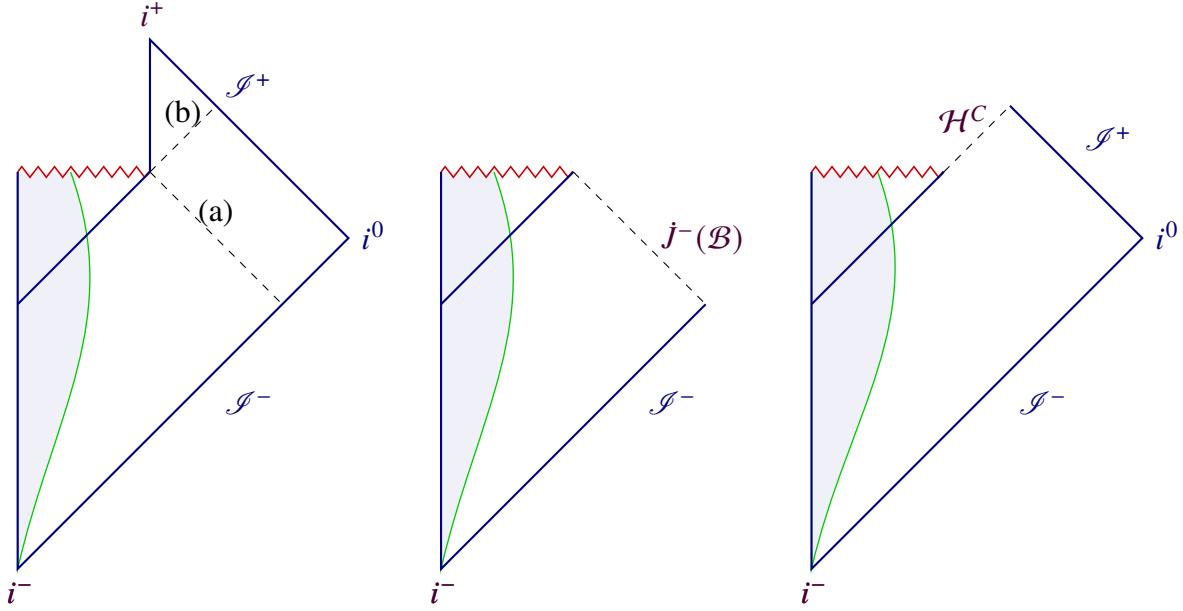
\begin{figure}
\centering
   \begin{tikzpicture}[scale=3.5]
  \message{Both embeddings}
 \coordinate (Oa) at (-1, 0); 
  \coordinate (NEa)  at (-0.5, 0.5); 
  \coordinate (NNa) at (-0.5, 1); 
  \coordinate (Sa)  at ( -1,-1); 
  \coordinate (Na)  at ( -1, 0.5); 
  \coordinate (Ea)  at ( 0.25, 0.25); 
  \coordinate (Xa)  at ( -0.8, 0.5); 
  \coordinate (Y1a) at (-0.25,0.75); 
  \coordinate (Y2a) at (0, 0);
  
  \draw[particle, fill = mylightblue]
      (Na) to (Sa) to[out=77,in=-70] (Xa);
  
  \draw[singularity] (Na) -- node[above] {} (NEa);
  
  \draw[thick,mydarkblue] (NEa) -- (NNa) -- (Ea) -- (Sa) -- (Na);
  \draw[thick,mydarkblue] (Oa) -- (NEa);

    \path (Oa) -- (NEa) node[mydarkblue,pos=0.75,below=4,scale=0.85]
    {\contour{mylightpurple}{}};

  \node[above=1,right=1,mydarkblue] at (Ea) {$i^0$};
  \node[right=1,below=1,mydarkpurple] at (Sa) {$i^-$};
  \node[right=1,above=1,mydarkpurple] at (NNa) {$i^+$};
  \node[right=1,below=1,mydarkpurple] at (Sa) {$i^-$};

  \node[mydarkblue,above right=-1] at (-0.25,0.75) {$\If$};
  \node[mydarkblue,below right=-1] at (-0.25,-0.3) {$\Ip$};

\tikzset{mylabel/.style  args={at #1 #2  with #3}{
    postaction={decorate,
    decoration={
      markings,
      mark= at position #1
      with  \node [#2] {#3};
 } } } }

    \draw[dashed, mylabel=at 0.5 above  with {(b)}]
      (NEa) to  (Y1a) ;

      \draw[dashed, mylabel=at 0.5 above with {(a)}]
      (NEa) to  (Y2a) ;

\message{causal past}

  \coordinate (Ob) at (0.6, 0); 
  \coordinate (NEb)  at (1.1, 0.5); 
  \coordinate (NNb) at (1.1, 1); 
  \coordinate (Sb)  at ( 0.6,-1); 
  \coordinate (Nb)  at ( 0.6, 0.5); 
  \coordinate (Eb)  at ( 1.85, 0.25); 
  \coordinate (Xb)  at ( 0.8, 0.5); 
  \coordinate (Y1b) at (1.35,0.75); 
  \coordinate (Y2b) at (1.6, 0);
  
  \draw[particle, fill = mylightblue]
      (Nb) to (Sb) to[out=77,in=-70] (Xb);
 
  \draw[singularity] (Nb) -- node[above] {} (NEb);
  
  \draw[thick,mydarkblue] (Y2b) -- (Sb) -- (Nb);
  \draw[thick,mydarkblue] (Ob) -- (NEb);

    \path (Ob) -- (NEb) node[mydarkblue,pos=0.75,below=4,scale=0.85]
    {\contour{mylightpurple}{}};
 
  \node[right=1,below=1,mydarkpurple] at (Sb) {$i^-$};
  \node[right=1,below=1,mydarkpurple] at (Sb) {$i^-$};
  \node[left = 6,mydarkpurple] at (Eb) {$\dot{J}^-(\BH)$};
  \node[mydarkblue,below right=-1] at (1.35,-0.3) {$\Ip$};

\tikzset{mylabel/.style  args={at #1 #2  with #3}{
    postaction={decorate,
    decoration={
      markings,
      mark= at position #1
      with  \node [#2] {#3};
 } } } }
  
      \draw[dashed]
      (NEb) to  (Y2b) ;

\message{MGHD scri minus}

    \coordinate (Oc) at (2, 0); 
  \coordinate (NEc)  at (2.5, 0.5); 
  \coordinate (NNc) at (2.5, 1); 
  \coordinate (Sc)  at ( 2,-1); 
  \coordinate (Nc)  at ( 2, 0.5); 
  \coordinate (Ec)  at ( 3.25, 0.25); 
  \coordinate (Xc)  at ( 2.25, 0.5); 
  \coordinate (Y1c) at (2.75,0.75); 
  
  \draw[particle, fill = mylightblue]
      (Nc) to (Sc) to[out=77,in=-70] (Xc);
  
  \draw[singularity] (Nc) -- node[above] {} (NEc);
  
  \draw[thick,mydarkblue] (Y1c) -- (Ec) -- (Sc) -- (Nc);
  \draw[thick,mydarkblue] (Oc) -- (NEc);

    \path (Oc) -- (NEc) node[mydarkblue,pos=0.75,below=4,scale=0.85]
    {\contour{mylightpurple}{}};

  \node[above=1,right=1,mydarkblue] at (Ec) {$i^0$};
  \node[right=1,below=1,mydarkpurple] at (Sc) {$i^-$};
  \node[right=1,below=1,mydarkpurple] at (Sc) {$i^-$};
  \node[above=20, right=-6,mydarkpurple] at (NEc) {$\mathcal{H}^C$};

  \node[mydarkblue,above right=-1] at (3,0.55) {$\If$};
  \node[mydarkblue,below right=-1] at (2.75,-0.3) {$\Ip$};

\tikzset{mylabel/.style  args={at #1 #2  with #3}{
    postaction={decorate,
    decoration={
      markings,
      mark= at position #1
      with  \node [#2] {#3};
 } } } }
  
    \draw[dashed]
      (NEc) to  (Y1c) ;
  
\end{tikzpicture}
    \caption{Left: The two globally hyperbolic regions of evaporation-Schwarzschild. The region below (a) is the causal past of the black hole, and the region below (b) is the MGHD of $\Ip$. Centre: The casual past of the black hole, $J^-(\BH)$. Right: The MGHD of $\Ip$, $D(\Ip)$.}
    \label{fig:Embeddings}
\end{figure}

Our first task is to find the region of evaporation-Schwarzschild in which to attempt to recover Hawking radiation. In quantum field theory on curved spacetimes, global hyperbolicity is nearly always assumed because this guarantees an initial value problem in the following sense: given an initial data surface in GR, there exists a unique (up to isometry) spacetime that is the maximal globally hyperbolic development (MGHD) of the data surface.
The initial data surface will be a Cauchy surface for this spacetime, and determines the entire spacetime. Moreover, it is clear that derivations of Hawking radiation which map modes in the past to modes in the future (as Hawking's and Fredenhagen and Haag's do) will require global hyperbolicity. This is because the state of the field in the past must determine the state of the field in the future. However, evaporation-Schwarzschild is not globally hyperbolic. Therefore, we must find a region of evaporation-Schwarzschild which is globally hyperbolic and has sufficient spacetime structure to admit Hawking's derivation of Hawking radiation. There are two reasonable sub-spacetime regions of evaporation-Schwarzschild which are globally hyperbolic: a) the causal past of the black hole region ($J^-(\BH)$), or b) the MGHD of $\Ip$, $(D(\Ip))$. These two embedded regions are demarcated in figure \ref{fig:Embeddings}.

Which is more suited to deriving Hawking radiation? It is MGHD $\Ip$. To see this, consider the causal past of the black hole spacetime. It is a spacetime such that, if a light ray were admitted at a point, it could reach the black hole before it evaporates completely. Near the evaporation event this is a tiny area, so we have deleted most of the spacetime we need for the derivation. 

More technically, `future null infinity' in the causal past of the black hole will be the boundary of the causal past, $\dot{J}^-(\BH)$. There will not be a timelike Killing vector field on this boundary; therefore, there will be no preferred time parameter with respect to which we can define a particle interpretation. This is because the boundary bisects the non-stationary exterior region of the spacetime. This also means the boundary won't be asymptotically flat; instead, it ends at the naked singularity and so it will contain a region of arbitrarily large curvature. Clearly, the causal past of the black hole region is useless for deriving Hawking radiation. 

MGHD $\Ip$ on the other hand does not suffer these problems, and includes the portion of $\If$ where all Hawking radiation will propagate to. Therefore this is the appropriate globally hyperbolic spacetime region to use.\footnote{It can be shown that neither the causal past of the black hole nor MGHD $\Ip$ is conformally equivalent to collapse-Schwarzschild, so proofs of the conformal equivalence of the Hawking temperature (e.g. \citeA{jacobson1993conformal}) do not help resolve the paradox.} So the question of this section is precisely stated as: which of the necessary assumptions for Hawking's derivation of Hawking radiation cannot be carried over into MGHD $\Ip$?

There are two important differences between MGHD $\Ip$ and collapse-Schwarzschild: the exterior solution is not Schwarzschild and the spacetime is not stationary. How do these changes affect the derivation? Firstly, Hawking's derivation makes use of ingoing and outgoing Eddington-Finkelstein coordinates, defined in equation \eqref{eq:EFcoords}, which are specified for a particular mass $m$. This constant mass term is unavailable in evaporation spacetimes. Instead, one must analyse how modes defined with respect to coordinates that cover MGHD $\Ip$ behave on an evaporation metric, but nothing like this is carried out for Hawking's derivation. 

Next, to calculate the form of modes on $\Ip$, Hawking exploits an isometry with the maximally extended Schwarzschild solution, and analyses modes that propagate onto the past horizon (see footnote \ref{fn:Maximal_isometry}). When the exterior solution is no longer Schwarzschild this isometry can not be used. Furthermore, and perhaps most strikingly, the failure of stationarity implies that the propagation of the modes back along the horizon will induce an evolution of the modes different to that calculated in collapse-Schwarzschild. Indeed, the normal null vector on the horizon which is used to compute the backwards evolution of the modes will have a different form in MGHD $\Ip$ as compared with collapse-Schwarzschild, precisely because the metric is different and the horizon area is changing. Finally, the non-stationarity will affect the scattering of the modes by the gravitational field.

Admittedly, the model of evaporation used here, evaporation-Schwarzschild, is heuristic only and not generally believed to be a realistic model of black hole evaporation. One may wonder whether in more realistic models of black hole evaporation the problems listed here go away. It is in fact the opposite, things are worse in realistic models. For example, in explicitly computed models \citeA{Schindler_Aguirre_Kuttner_2020} show that, as well as the above worries still holding true, there is also no event horizon or Killing horizon for an evaporating black hole. Thus there will be no null vector normal to the horizon at all; the very structure Hawking uses to compute the form of the modes on $\Ip$ is non-existent in evaporation spacetimes. So, in realistic evaporation models, more of the spacetime structure exploited by Hawking to derive the radiation is lost. 

Hawking himself notes that the ``negative energy flux will cause the area of the event horizon to decrease and so the black hole will not, in fact, be in a stationary state'' \citeyear[p. 219]{Hawking:1975vcx}. He accepts this is a problem, but claims to have a solution, as one can approximate the black hole as ``quasi stationary''. In section \ref{ssec:approx_regime} I show that this approximation regime does not, in fact, save Hawking's derivation because one cannot use the regime to recover the necessary global structure. Therefore, the problems remain.\footnote{One interpretation of what's at stake here is the thermality of Hawking radiation, which dovetails with similar arguments associated with the information paradox. A serious discussion of this point would require an extended analysis. I thank an anonymous referee for emphasising this aspect of the problem.}

A reader familiar with the vast literature of derivations of Hawking radiation may at this point be thinking of their preferred derivations, and be under the impression that they do not fall victims to the above challenges. I have no objection to such claims. Indeed I will present certain derivations as the best candidates currently available to resolve the paradox for Hawking's derivation in section \ref{ssec:Implicit_structure}. Nevertheless, this is not a problem for my thesis as I am focused on particular derivations of Hawking radiation, in this case Hawking's original derivation. Thus, given the amount of structure exploited by Hawking which does not carry over to MGHD $\Ip$, one must accept the conclusion that the derivation of Hawking radiation found in \citeA{Hawking:1975vcx} falls victim to the idealization paradox. That is to say, remarkably, Hawking's derivation is evaporation-inconsistent!

\citeA{Fredenhagen:1989kr} construct their derivation to avoid a different problematic assumption in Hawking's derivation, the geometric optics approximation. Thus, Fredenhagen and Haag's derivation is what most consider to be the watertight derivation. It is to this that I turn next.

\section{Idealization Paradox in Fredenhagen and Haag's derivation}\label{sec: FH_paradox}

Fredenhagen and Haag's derivation is similar to Hawking's in that it defines the state outside the black hole at some early time and maps this state to some state at late time. However, it differs in a few important respects. Firstly, the entire calculation is performed on the region of spacetime after the stellar matter has passed the event horizon. Secondly, they use a `detector' at asymptotically late times to model the radiation. Thirdly, they perform the calculation by propagating the detector along the timelike Killing vector field in the exterior region. I sketch this derivation next, and in section \ref{ssec:FH_Failure} show that it is also evaporation-inconsistent.

\subsection{Sketch of Fredenhagen and Haag's Derivation}\label{ssec:FH_sketch}

This derivation, like Hawking's, takes place on collapse-Schwarzschild. The region exterior to the event horizon in Schwarzschild can be covered by the coordinates $(t,r, \theta, \phi)$, where we call $t$ Schwarzschild-time, and define $\tau$-time coordinates, $(\tau, r, \theta, \phi)$ where $\tau = t+r^*-r = v -r$, for $v$ and $r^*$ defined in \eqref{eq:EFcoords}. $\tau$ is approximately Schwarzschild-time near spacelike infinity, and becomes infinitely negative near the horizon. Let $r = r_s(\tau)$ define the surface of the collapsing star, with $r_s(0)=r_0$ the Schwarzschild radius, such that the star crosses the Schwarzschild radius at $\tau=0$. 

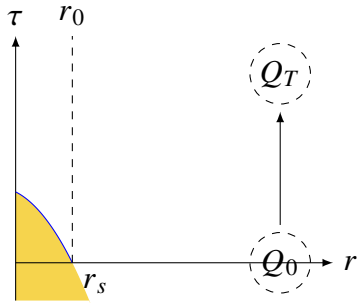
\begin{figure}[tbh]\begin{center}
    \begin{center}

  \begin{tikzpicture}
   \fill[myorange!70]  plot[domain=0:1]  ({\x}, {-(\x+0.25)^2+1})  --
 (0.97,-0.5) -- (0, -0.5) -- cycle;
  \draw[->] (0, 0) -- (4.2, 0) node[right] {$r$};
  \draw[->] (0, -0.5) -- (0, 3) node[above] {$\tau$};
  \draw[domain=0:0.75, smooth, variable=\x, blue] plot ({\x}, {-(\x+0.25)^2+1}) node[below right, black ] {$r_s$};

 \draw[dashed]{} (3.5, 0) circle (0.4) node {$Q_0$};
 \draw[dashed]{} (3.5, 2.5) circle (0.4) node {$Q_T$};
 \draw[->] (3.5, 0.5) -- (3.5, 2);
 \draw[dashed] (0.75, 0) -- (0.75, 3) node[above] {$r_0$};
 
\end{tikzpicture}
\end{center}  
    
  \end{center}
  \caption{Set up of Fredenhagen and Haag's derivation, in which a detector $Q$ is propagated along the timelike Killing vector field to asymptotically late times, $Q_T(T\rightarrow\infty)$.}\label{fig:FH_tau-time}\end{figure}

As before, let $\Phi$ be a massless complex-valued scalar quantum field which satisfies the covariant wave equation. Fredenhagen and Haag model a detector in a spacetime region $O$ with an observable, $Q^*Q$, which is the counting rate given by $\langle Q^*Q \rangle$, where $Q = \int \Phi(x)h(x)\sqrt{|g|}d^4x$ for a test function $h(x)$ that has support in $O$. They `place' the detector at a large radius at the $\tau$-time for which the collapsing star crosses the horizon (that is, $h$ has support around  $(0, R, \theta_0, \phi_0)$ for $R \gg r_0$). The detector is then translated along the timelike Killing vector field of the Schwarzschild metric. We are interested in the counting rate of the detector at asymptotically late times (given by $Q_T(T\rightarrow \infty)$), as displayed in figure \ref{fig:FH_tau-time}, in which a collapsing star and the time-translated detector are displayed in $\tau$-time coordinates. 

\begin{figure}
    \centering
    \begin{tikzpicture}[scale=2.3]
  \message{Extended Penrose diagram: Schwarzschild black hole^^J}
  
  \def\R{0.08} 
  \def\Nlines{3} 
  \pgfmathsetmacro\ta{1/sin(90*1/(\Nlines+1))} 
  \pgfmathsetmacro\tb{sin(90*2/(\Nlines+1))}   
  \pgfmathsetmacro\tc{1/sin(90*2/(\Nlines+1))} 
  \pgfmathsetmacro\td{sin(90*1/(\Nlines+1))}   
  \coordinate (-O) at (-1, 0); 
  \coordinate (-S) at (-1,-1); 
  \coordinate (-N) at (-1, 1); 
  \coordinate (-W) at (-2, 0); 
  \coordinate (-E) at ( 0, 0); 
  \coordinate (O)  at ( 1, 0); 
  \coordinate (S)  at ( 1,-1); 
  \coordinate (N)  at ( 1, 1); 
  \coordinate (E)  at ( 2, 0); 
  \coordinate (W)  at ( 0, 0); 
  \coordinate (B)  at ( 0,-1); 
  \coordinate (T)  at ( 0, 1); 
  \coordinate (X0) at (-0.4,1);
  \coordinate (X1) at ({asin(sqrt((\tc^2-1)/(\tc^2-\td^2)))/90},
                       {acos(\tc*sqrt((1-\td^2)/(\tc^2-\td^2)))/90}); 
  \coordinate (X2) at (45:0.87); 
  \coordinate (X3) at (0.60,1.05); 
  
   \fill[myorange!70]{}
  (S) to[out=100, in =-40] (X0) to (-N) to (-W) to (-S) to (S);
  
  \draw[singularity] (-N) -- node[above] {future singularity $r=0$} (N);
  \draw[singularity] (S) -- node[below] {past singularity $r=0$} (-S);
  \path (S) -- (W) node[mydarkblue,pos=0.50,below=-2.5,rotate=-45,scale=0.85]
    {};
  \path (W) -- (N) node[mydarkblue,pos=0.50,above=-2.5,rotate=45,scale=0.85]
    {};
  \draw[thick,mydarkblue] (-N) -- (-E) -- (-S) -- (-W) -- cycle;
  \draw[thick,mydarkblue] (N) -- (E) -- (S) -- (W) -- cycle;

  \node[above=1,left=1,mydarkblue] at (-2,0) {$i^0$};
  \node[above=1,right=1,mydarkblue] at (2,0) {$i^0$};
  \node[right=1,below=1,mydarkpurple] at (-S) {$i^-$};
  \node[right=1,above=1,mydarkpurple] at (-N) {$i^+$};
  \node[right=1,below=1,mydarkpurple] at (S) {$i^-$};
  \node[right=1,above=1,mydarkpurple] at (N) {$i^+$};
  \node[mydarkblue,below left=-1] at (-1.5,-0.5) {$\calI^-$};
  \node[mydarkblue,above left=-1] at (-1.5,0.5) {$\calI^+$};
  \node[mydarkblue,above right=-1] at (1.5,0.5) {$\calI^+$};
  \node[mydarkblue,below right=-1] at (1.5,-0.5) {$\calI^-$};
  
   \draw[dashed]{} (1, 0.6) circle (0.071) node[scale=0.5] {$Q_T$};
   \draw[->, thick] (1, 0.7) -- (1, 0.95);
   \draw[dashed] (1.1, 0.6) -- (1.86, -1.86+1.7);
   \draw[dashed] (0.95, 0.55) -- (1.75, -1.75+1.5);
   \draw[dashed] (1.05, 0.55) -- (0.25, 0.25-0.5);
   \draw[dashed] (0.95, 0.65) -- (0.16, 0.16-0.3);

   \fill[particle]
  (1.75, -1.75+1.5) to[out=45, in=225] (1.75, -1.75+1.6) to[out=45, in=225] (1.86, -1.86+1.7);

  \fill[particle]
  (0.16, 0.16-0.3) to[out=-45, in=-225] (0.24, 0.16-0.3) to[out=-45, in=-225] (0.25, 0.25-0.5);

  \draw[->, thick]
  (0.2, 0.25-0.5) to (0, 0.45-0.5);

  \draw[->, thick]

  (1.8, -1.75+1.5) to (2.0, -1.55+1.5);

  \draw[world line]
  (E) to[out = 163, in=17] (-W);
\end{tikzpicture}
   \caption{Decomposition of modes contributing to detector response of maximally extended Schwarzschild.}\label{fig:Wald_on_FH}
\end{figure}
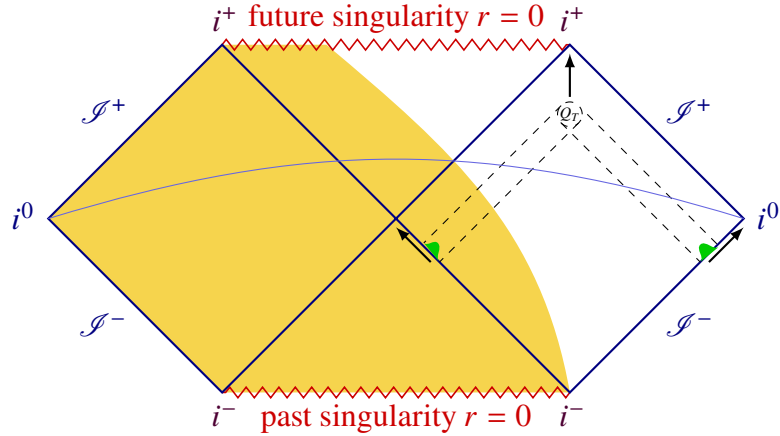

The counting rate is determined by the data on a Cauchy surface in the past of the late time detector. In the asymptotic limit, $T \rightarrow \infty$, the contributing data on the Cauchy surface decomposes into a sum of two wave packets, shifting asymptotically to $r \rightarrow \infty $ and $r \rightarrow r_0$. \citeA[pp. 159-162]{Wald:1995yp} explains this fact by noting that in maximally extended Schwarzschild, any mode in the region exterior to the black hole will decompose into modes on the past horizon and $\Ip$. Propagating this decomposition along Killing vector fields infinitely far will place modes infinitely close to the future horizon and spatial infinity, as depicted in figure \ref{fig:Wald_on_FH}. By isometry, we can draw the same conclusion for the exterior of collapse-Schwarzschild.\footnote{This inference is not in fact secure because the MGHD of the spacetime region exterior to the collapsing matter is not maximally extended Schwarzschild, but I ignore this difficulty here as it does not undermine Fredenhagen and Haag's calculation but only Wald's explanation of the behaviour of the wave-packet decomposition.} This fact can also be seen as a consequence of the Schwarzschild potential pushing modes onto the horizon and out to infinity. 

Assuming the state in the distant past is vacuum, the contribution to the counting rate from spatial infinity is zero. The contribution from the wave-packet that accumulates at the horizon is determined by the short-distance behaviour of the quantum field. Without concerning ourselves with the details, the authors assume the leading singularity in the short distance behaviour has a particular form that is implied by the Hadamard condition. Armed with this assumption, Fredenhagen and Haag show that the modes on the horizon contribute a thermal spectrum to the counting rate of the detector at asymptotically late times, with temperature given by the Hawking temperature.

\subsection{Fredenhagen and Haag's Derivation Fails in Evaporation-Schwarzschild}\label{ssec:FH_Failure}\label{ssec:FH_Failure}

We are again interested in whether we can deidealize this derivation: can the assumptions necessary to carry out Fredenhagen and Haag's derivation be carried over to MGHD $\Ip$? 

The most prominent difficulty for the derivation is that it relies on the stationarity of the exterior metric. The detector is time translated along the Killing vector field of the Schwarzschild metric. This sends the modes on the Cauchy surface that contribute to the counting rate to spatial infinity and onto the horizon. Moreover, the behaviour of the mode decomposition as the detector is time translated is analysed on the maximally extended Schwarzschild spacetime and, following Wald, arises due to the global Killing field that is timelike in the exterior region. 

In MGHD $\Ip$ there are no such timelike Killing vector fields and there is no isometry with maximally extended Schwarzschild because the size of the black hole is changing. Indeed, there do not even exist approximate Killing vector fields on the entire spacetime, whatever notion of `approximate' one might try to use. MGHD $\Ip$ contains a large mass black hole at $\tau=0$, and by the evaporation event it contains a negligible mass black hole. This is clearly a radical change and so the spacetime is in no sense stationary. The behaviour of modes under the time translation symmetry of collapse-Schwarzschild was the core of the derivation, and this is simply not available in evaporation-Schwarzschild. 

In addition, there is a further difference between the two spacetimes relevant to Fredenhagen and Haag. The global time function on MGHD $\Ip$ does not extend to infinity into the future, whereas it is future infinite on collapse-Schwarzschild. The lack of a future-infinite time coordinate is a problem for Fredenhagen and Haag's derivation because, whereas Hawking's asymptotic time assumption was realised by future null infinity, Fredenhagen and Haag translate their detector along a timelike worldline. Every timelike worldline will reach the Cauchy horizon of MGHD $\Ip$ in a finite parameter distance, so one cannot take the asymptotic time limit. This limit was essential to Fredenhagen and Haag's derivation as it pushed the modes asymptotically close to the horizon, forcing them into the trans-Planckian regime. Fredenhagen and Haag can then describe the modes by their short distance behaviour. Without this limit, we cannot be sure of the derivation.

The Fredenhagen and Haag derivation cannot be carried out, in any obvious fashion, in evaporation-Schwarzschild. The idealization paradox thus applies to this approach as well: it too is evaporation-inconsistent. I now turn to algebraic approaches. 


\section{Idealization Paradox in Algebraic Approaches}\label{sec:Algebraic_Approach}

\subsection{Sketch of the Algebraic Derivation}\label{ssec:Algebraic_sketch}

I will not go into any sort of detail in the sketch of algebraic approaches, as they are on the one hand very mathematically heavy, but on the other very conceptually simple. Algebraic approaches function by showing a particular state is the uniquely natural stationary Hadamard vacuum state on the collapse-Schwarzschild spacetime, and that this state is thermal at the Hawking temperature at future null infinity.

Algebraic QFT begins with a $*$-algebra of observables $\mathcal{A}$. A state $\omega$ is a completely positive map from $\mathcal{A}$ to $\mathbb{C}$, $\omega: \mathcal{A} \rightarrow \mathbb{C}$. For self-adjoint operators the map is real valued. We fix states by demanding they obey certain conditions, such as being vacuum.\footnote{In curved spacetimes, the algebraic approach defines a vacuum as a state that is Gaussian and pure (see \citeA{KAY199149} for details).} Conversely, we can discover facts about states by assessing what conditions they obey, for example a state is thermal with respect to a given Hamiltonian if it obeys the KMS condition.\footnote{See \citeA[p. 13]{bratelli1982operator} for a definition.} As usual, we demand that physical states are Hadamard. Finally, one can define the algebra of observables for a scalar field by demanding that the functions used to smear the observables solve the covariant-wave equation. 

One finds (\citeA{DIMOCK1987366}, \shortciteA{dappiaggi2011rigorous}) that the uniquely natural stationary Hadamard vacuum state on the collapse-Schwarzschild spacetime is the Unruh vacuum. The Unruh vacuum has the property of having no particles near $\Ip$, but being thermal at the Hawking temperature near $\If$, with a flux going out to infinity. Thus, one claims that the black hole is emitting Hawking radiation.

This sketch is sufficient to analyse the idealization paradox for algebraic approaches, to which I turn now.

\subsection{Algebraic Approaches Fail in Evaporation-Schwarzschild}\label{ssec:A_Failure}


Algebraic approaches are the most mathematically rigorous formulation of Hawking radiation. However, they clearly fail to survive the move to evaporation-Schwarzschild, or MGHD $\Ip$. 

The spacetime we are now interested in is not collapse-Schwarzschild, and not even approximately collapse-Schwarzschild. Therefore, the proof of the unique naturalness of the Unruh vacuum simply does not apply; the Unruh vacuum is uniquely natural on collapse-Schwarzschild, with no implication for the uniquely natural vacuum state on MGHD $\Ip$. Moreover, given that one condition on the Unruh vacuum is that it is stationary, and collapse-Schwarzschild is not stationary, clearly the Unruh vacuum will be the inappropriate vacuum state for MGHD $\Ip$. We can thus conclude that the idealization paradox applies to the algebraic approaches. 

To conclude, we have three different derivations, each of increasing mathematical rigour, and each with open questions about how they can actually claim to be establishing Hawking radiation in physically realistic models.

\section{Paths Toward a Resolution}\label{sec:Discussion}

Physics uses idealizations all the time. The idealization used in the derivations here is only particularly striking because it leads to a paradox, rendering the argument each derivation presents for Hawking radiation inconsistent. This paradox is clearly unacceptable, and so we should find a resolution. One aim of this paper, the aim taken up in this section, is to categorise and assess solutions to the idealization paradox. The most natural resolutions are those that deidealize the derivations, to show how they can proceed in evaporation spacetimes. Three sub-categories of deidealization solutions are presented below:

\begin{itemize}
    \item Quantum Gravity (section \ref{ssec:QuantumGravity})
    \item Approximation Regime (section \ref{ssec:approx_regime})
    \item Essential Structure (section \ref{ssec:Implicit_structure})
\end{itemize}
The first argues that quantum gravity is needed to describe black hole evaporation and thus resolve the paradox. The second looks to find an approximation regime between collapse-Schwarzschild and evaporation-Schwarzschild. Specifically, I formalise and analyse an approximation regime suggested in \citeA{Hawking:1975vcx}. The third argues that one can weaken the assumptions of the derivations, such that each derivation can derive Hawking radiation whilst assuming only some essential spacetime structure that is present in both evaporation and non-evaporation spacetimes.

I find that quantum gravity holds no prospects for resolving the paradox. I find Hawking's approximation regime achieves varying degrees of success for the different derivations, but even where there are hints of success more work is needed. Finally, I find that essential structure derivations constitute a very fruitful research direction which has already been taken up in \citeA{VISSER_2003, barcelo2011minimal,barcelo2011hawking}. Indeed, this work already points towards deep lessons about the nature of Hawking radiation.

The derivations analysed in this paper, as discussed in the introduction, are not exhaustive. So, plausibly, other derivations don't face the paradox (indeed I will discuss some examples in section \ref{ssec:Implicit_structure}), and such derivations are to be preferred assuming one seeks a consistent physical theory. Given a paradox-free derivation, the consistency conjecture will be true, and so the phenomenon of Hawking radiation will be insulated from the paradox.

However, this still leaves us with an idealization paradox for at least some derivations. The paradox still needs to be resolved for these derivations because there are important lessons available in at least three dimensions: 1) the relationship between evaporation-consistent and -inconsistent derivations will inform our physical interpretation of Hawking radiation; 2) different de-idealizations connote different understandings of the idealization used, for example: an essential structure deidealization (section \ref{ssec:Implicit_structure}) suggests a reduction and no idealized global structure, but an approximation regime (section \ref{ssec:approx_regime}) suggests no reduction and idealized global structure; 3) it is historically important how and when such a scientifically revolutionary piece of physics was rendered consistent.


I do not consider here resolutions which may be collected under the name deidealization pessimism, examples of such views include: embracing evaporation-inconsistent derivations as essential idealizations (aligning with \citeA{Batterman2002-BATTDI, BATTERMAN2005225, Batterman2011-BATESA}), and denying the phenomenon of either black hole evaporation or Hawking radiation. Such approaches would resolve the paradox, but offer a somewhat pyrrhic victory by respectively rejecting either Earman's principle, or the consensus in black hole physics.\footnote{\citeA{wallace2018case,WALLACE2019103} reviews the arguments in favour of this consensus.} Instead, the categories I propose below (in sections \ref{ssec:QuantumGravity}, \ref{ssec:approx_regime} and \ref{ssec:Implicit_structure}) help to distinguish different ways a derivation may be deidealized to avoid the paradox.

\subsection{Quantum Gravity}\label{ssec:QuantumGravity}

It is widely believed that a quantum theory of gravity will resolve the black hole information paradox.\footnote{For taxonomies of such proposals see \citeA{Belot1999, Unruh2017InformationL}.} This is because the consensus in the physics community is that a quantum theory of gravity will be required to describe the final stages of black hole evaporation (for example, \citeA{Rovelli_2014}). Moreover, it is often claimed that the early stages of black hole evaporation also cannot be fully described without a quantum theory of gravity, as we can't accurately describe the backreaction of Hawking radiation on the metric. Therefore, a reasonable first suggestion is to expect quantum gravity to resolve the idealization paradox. However, I argue this proposal cannot succeed. 

The central idea of a quantum gravity resolution to the idealization paradox is that the physics of spacetimes and Hawking radiation occurs in the semi-classical limit, whereas black hole evaporation occurs in a full quantum gravity description. One would argue that this allows one to reject the Inconsistency Claim, premise 3 of the idealization paradox presented in the introduction, which asserts that black hole evaporation leads to the rejection of assumptions required for the derivation of Hawking radiation. In order to reject the Inconsistency Claim, one may argue that because black hole evaporation is a quantum gravity phenomenon, it is not describable in the semi-classical limit and as such tells us nothing about the properties of spacetime in the semi-classical limit. Thus one cannot infer from evaporation to the breakdown of the semi-classical limit spacetime properties required for Hawking radiation. Thus, by acknowledging the need for a quantum theory of gravity to describe black hole evaporation, we can escape the paradox.

Unfortunately, quantum gravity does not license us to reject the Inconsistency Claim. To see this, note that any quantum gravity theory of black hole evaporation must be able to represent: i) a black hole of given mass-energy, and ii) the mass-energy of a black hole being reduced in the process of evaporation. If the mass-energy of a black hole is not reducing then one cannot claim to be describing black hole evaporation, it is some other phenomena. This is certainly a possibility, but such a theory would constitute evaporation scepticism by claiming Hawking radiation does not lead black holes to lose mass-energy.   

Given these minimal representational requirements, the state in our quantum theory of gravity will represent a black hole of mass $m_1$ in the semi-classical limit at some earlier time, and a black hole of mass $m_2$ in the semi-classical limit at some later time, where $m_1>m_2$. This immediately violates stationarity, one of the properties used in the derivations of Hawking radiation discussed here. Therefore, even a completely quantum gravity model of evaporation implies the breakdown of properties required for derivations of Hawking radiation in the semi-classical limit. Hence we are not licensed to reject the Inconsistency Claim. 

Why does the black hole information paradox admit a quantum gravity resolution whereas the idealization paradox does not? The difference is that there is no black hole information paradox until the evaporation event\footnote{In the traditional sense, though not in the Page-time paradox sense; see \citeA{wallace_2020}.} because only then does one have to accept the information has vanished from the universe. Moreover, there is no minimal representational requirement on the evaporation event so we cannot anticipate any aspect of the quantum gravity description. On the other hand, the idealization paradox arises without the need to consider the evaporation event, due to the failure of properties in the entire exterior region such as stationarity. We can then impose our minimal condition on evaporation far before the evaporation event, and this leads to the paradox.

\subsection{Approximation Regime}\label{ssec:approx_regime}

Perhaps the derivations considered here can be carried out in some appropriate approximation regime: One would find some spacetime region in collapse-Schwarzschild which looks approximately like some corresponding region of evaporation-Schwarzschild. One could then hope to carry out the derivation using this approximating region of collapse-Schwarzschild, then infer the derived radiation back onto evaporation-Schwarzschild. Thus, one would derive the existence of Hawking radiation in the evaporation spacetime. \citeA[p. 219]{Hawking:1975vcx} proposed such a resolution to the paradox: ``it is a reasonable approximation to describe the black hole by a sequence of stationary solutions and to calculate the rate of particle emission in each solution.'' 

The regime is justified as follows: The rate of change of the mass of the black hole will (for masses larger than the Planck mass) be much slower than the time taken for light to propagate to a region that can be modelled as approximately flat.\footnote{Initial work on modelling `infinity' at a finite distance so asymptotic flatness can be defined for realistic sub-systems of the Universe can be found in \citeA[sec. 5]{ELLIS2002645}. Such modelling frameworks will likely be helpful for rigorously analysing approximation regimes such as the one proposed here. However, finite-infinity models are tangential to our current concern because I accept the standard presumption of the literature that astrophysical black holes are well modelled by collapse-Schwarzschild. If one demands finite-infinity models, it is then necessary to show how derivations of Hawking radiation look in these deidealized models, and what the relationship of such finite-infinity models to the models discussed in this paper is.} Thus, one can approximate the variable mass black hole spacetime as a sequence of stationary regions and calculate the rate of particle emission in each solution, avoiding the non-stationarity issues.

This is a very intuitive picture if one imagines a black hole as a compact three-dimensional object that evolves in time and produces Hawking radiation via a local mechanism. However, as we have seen, the derivations of Hawking radiation discussed above use global spacetime structure, including the propagation of modes through the collapsing matter region \cite{Hawking:1975vcx} and timelike Killing vector fields with time parameter extended into the infinite future \cite{Fredenhagen:1989kr}. Consequently, we should not assume that the slow rate of evaporation is sufficient to guarantee the derivations are unaffected; indeed, to do so would be negligent of philosophers of physics seeking to understand the derivations, idealizations, and phenomena at hand. 

\begin{figure}
 \begin{minipage}{.2\textwidth}
    \begin{tikzpicture}[scale=2.5]
  
  \coordinate (O) at (0, 0); 
  \coordinate (NE)  at (0.5, 0.5); 
  \coordinate (NN) at (0.5, 1); 
  \coordinate (S)  at ( 0,-1); 
  \coordinate (N)  at ( 0, 0.5); 
  \coordinate (E)  at ( 1.25, 0.25); 
  \coordinate (X)  at ( 0.5, 0.35); 
  \coordinate (Y) at (0.5, 0.2); 
  \coordinate (Z) at (0.1, 0.5);

\clip[decorate,decoration={zigzag,amplitude=2,segment length=6.17}]
 (N) -- (NE) --++ (-0.5,1) --++ (1.5,0) |-++(-3,-3) |- (-0.5,1) -- cycle;

\filldraw[fill=black!30, decoration={snake,amplitude=0.9,segment length=4,post length=3.8}] (X) 
 [snake=snake, draw=red] -- ++ (0.32, 0.32) 
 [snake=none, draw=black] -- (0.9,0.6) 
 [snake=snake, draw=red] -- (Y)
  [snake=snake, draw=red] -- ++(-0.4,0.4)
 [snake=none, draw=black] -- (0.26, 0.59)
  [snake=snake, draw=red] -- (X);
  
  \draw[particle, fill = mylightblue]
      (N) to (S) to[out=77,in=-70] (Z);
  
  \draw[singularity] (N) -- node[above] {} (NE);
  
  \draw[thick,mydarkblue] (NE) -- (NN) -- (E) -- (S) -- (N);
  \draw[thick,mydarkblue] (O) -- (NE);

    \path (O) -- (NE) node[mydarkblue,pos=0.75,below=4,scale=0.85]
    {\contour{mylightpurple}{}};

  \node[above=0,left=1,mydarkblue] at (O) {$r=0$};
  \node[above=1,right=1,mydarkblue] at (E) {$i^0$};
  \node[right=1,below=1,mydarkpurple] at (S) {$i^-$};
  \node[right=1,above=1,mydarkpurple] at (NN) {$i^+$};
  \node[right=1,below=1,mydarkpurple] at (S) {$i^-$};

  \node[mydarkblue,above right=-1] at (1.1,0.4) {$\If$};
  \node[mydarkblue,below right=-1] at (1.05,0.1) {$\Ip$};

\clip[decorate,decoration={zigzag,amplitude=2,segment length=6.17}]
 (N) -- (NE) --++ (-0.5,1) --++ (1.5,0) |-++(-3,-3) |- (-0.5,1) -- cycle;



    \path ( 0.82, 0.65) node[above right, black] (A)  {$m-\Delta m$};
    \path (0.9,0.55) node[above right, black] (A)  {$m$};

     \draw[dashed]{} (NE) -- ++ (0.23,0.25)  node[above right, black] (A)  {$m=0$};

\draw[-{Triangle[length=2mm,width=2.5mm]}, red] (0.85,0.55) --++ (0.05,0.05);
\draw[-{Triangle[length=2mm,width=2.5mm]}, red] (0.77, 0.62) --++ (0.05,0.05);
\draw[-{Triangle[length=2mm,width=2.5mm]}, red] (0.37,0.48) --++ (-0.05,0.05);
\draw[-{Triangle[length=2mm,width=2.5mm]}, red] (0.22, 0.47) --++ (-0.05,0.05); 
\end{tikzpicture}
\end{minipage}
\begin{minipage}{.2\textwidth}
\begin{tikzpicture}[scale=2.5]
  \message{collapse-Schwarzschild}
  
  \coordinate (O) at (0, 0); 
  \coordinate (NE)  at ( 1, 1); 
  \coordinate (S)  at ( 0,-1); 
  \coordinate (N)  at ( 0, 1); 
  \coordinate (E)  at ( 1.5, 0.5); 
  \coordinate (X)  at ( 0.2, 1); 
  \coordinate (Y) at (0, -0.3); 
  \coordinate (A) at (1, 0.8);
  \coordinate (B) at (1, 0.6);
  
  \node[above=0,left=1,mydarkblue] at (O) {$r=0$};
  \node[above=1,right=1,mydarkblue] at (E) {$i^0$};
  \node[right=1,below=1,mydarkpurple] at (S) {$i^-$};
  \node[right=4,above=1,mydarkpurple] at (NE) {$i^+$};
  \node[right=1,below=1,mydarkpurple] at (S) {$i^-$};

\clip[decorate,decoration={zigzag,amplitude=2,segment length=6.17}]
      (N) -- (NE) |-++ (0.75,0.3) |-++ (-4,-2.5) |- cycle;
\tikzset{mylabel/.style  args={at #1 #2  with #3}{
    postaction={decorate,
    decoration={
      markings,
      mark= at position #1
      with  \node [#2] {#3};
 } } } }

 \filldraw[fill=black!30, decoration={snake,amplitude=0.9,segment length=4,post length=3.8}, ] (A) 
 [snake=snake, draw=red] -- ++ (0.1, 0.1)
 [snake=none, draw=black] -- (1.2, 0.8)
 [snake=snake, draw=red] -- (B)
  [snake=snake, draw=red] -- ++(-0.45,0.45)
 [snake=none, draw=black] -- (0.6, 1.2)
  [snake=snake, draw=red] -- (A);
  
  \draw[particle, fill = mylightblue]
      (N) to (S) to[out=77,in=-70] (X);
  
  \draw[singularity] (N) -- node[above] {} (NE);
  
  \draw[thick,mydarkblue] (NE) -- (E) -- (S) -- (N);
  \draw[thick,mydarkblue] (O) -- (NE);

    \path (O) -- (NE) node[mydarkblue,pos=0.75,below=4,scale=0.85]
    {\contour{mylightpurple}{}};

  \node[mydarkblue,above right=-1] at (1.25,0.75) {$\If$};
  \node[mydarkblue,below right=-1] at (0.75,-0.3) {$\Ip$};



\draw[-{Triangle[length=2mm,width=2.5mm]}, red] (1.15,0.75) --++ (0.05,0.05);
\draw[-{Triangle[length=2mm,width=2.5mm]}, red] (1.05,0.85) --++ (0.05,0.05);
\draw[-{Triangle[length=2mm,width=2.5mm]}, red] (0.82,0.97) --++ (-0.05,0.05);
\draw[-{Triangle[length=2mm,width=2.5mm]}, red] (0.63,0.97) --++ (-0.05,0.05);
\end{tikzpicture}
\end{minipage}
 \caption{The two shaded regions are approximately isometric.}\label{fig:Quasi-stationary}
\end{figure}
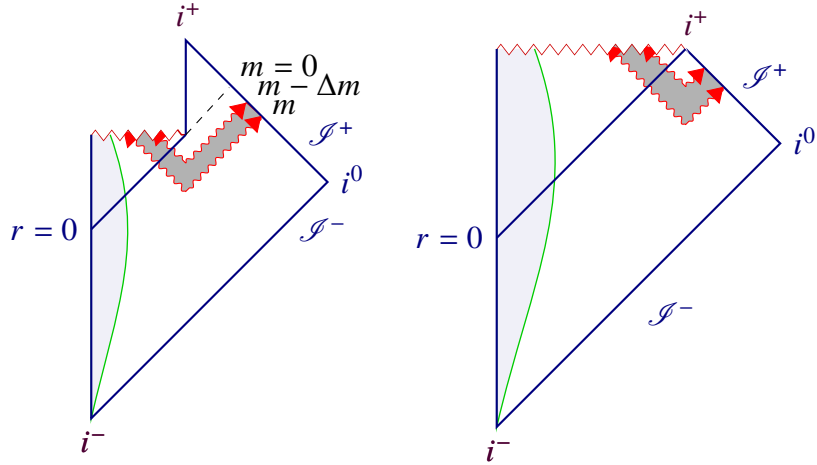

To give Hawking a more charitable treatment, let me propose a more promising way to formalise this approximation. We want to identify regions of evaporation-Schwarzschild which are approximated by regions of collapse-Schwarzschild so that the derivations can be carried out using collapse-Scwharzschild. The best candidate regions are the parts of spacetime which Hawking radiation propagates through as it escapes to future null infinity. Figure \ref{fig:Quasi-stationary} illustrates the structure of the regions (following \citeA[p. 178]{Wald:1995yp}). We model photons carrying energy away from the black hole to $\If$ and a negative energy flux propagating over the horizon. This is symbolised by two red arrows emerging from a single point, one pointing over the event horizon, the other out to future null infinity. Two such photon emission events are displayed in each conformal diagram in figure \ref{fig:Quasi-stationary}. The shaded region between the two photon emission events in evaporation-Schwarzschild is quasi-stationary. Thus the corresponding shaded region of the collapse-Schwarzschild spacetime of mass $m$ is approximately isometric to the shaded region of evaporation-Schwarzschild, where $m$ is the mass of the black hole according to the quasi-stationary region.\footnote{The mass of the slices will have to be modelled by the Bondi mass, as this is defined at future null infinity whereas the ADM mass can only be defined at spacelike infinity.}

I denote such a quasi-stationary region of evaporation-Schwarzschild as $\mathcal{R}_{QS}$ and a corresponding approximately isometric stationary region of a collapse-Schwarzschild spacetime as $\mathcal{R}_{S}$. The regime will work by using $\mathcal{R}_{S}$ instead of $\mathcal{R}_{QS}$ to derive Hawking radiation. One then infers the same result, to some degree of approximation, in the approximately isometric $\mathcal{R}_{QS}$. Repeating this for every $\mathcal{R}_{QS}$ should describe the Hawking effect in evaporation-Schwarzschild. 

This regime will face two central problems. First, for each of the derivations $\mathcal{R}_{S}$ will have insufficient structure to derive Hawking radiation because it is a smaller extendable subspacetime. Thus it will be necessary to use the MGHD of $\mathcal{R}_{S}$. Problematically, although $\mathcal{R}_{QS}$ and $\mathcal{R}_{S}$ are approximately isometric, the corresponding MGHD for each will be very different. So although it is clear we can use $\mathcal{R}_{S}$ to draw approximately correct conclusions about $\mathcal{R}_{QS}$, it is less clear that we can use the MGHD of $\mathcal{R}_{S}$ to draw approximately correct conclusions about $\mathcal{R}_{QS}$. One must therefore justify using the MGHD of $\mathcal{R}_{S}$ rather than only the approximately isometric region, $\mathcal{R}_{S}$, despite the different global structure. 

Second, even if one can justify using the MGHD of $\mathcal{R}_{S}$ to draw inferences about $\mathcal{R}_{QS}$, neither $\mathcal{R}_{QS}$ nor $\mathcal{R}_{S}$ contain Cauchy surfaces for evaporation-Schwarzschild or collapse-Schwarzschild respectively. This is obvious in figure \ref{fig:Quasi-stationary} where, for example, a massive particle can travel from $i^-$ and reach the singularity and never record data on the quasi-stationary surface. The same is true for $\mathcal{R}_{S}$ in collapse-Schwarzschild. This means that neither $\mathcal{R}_{QS}$ nor its approximately isometric stationary sibling $\mathcal{R}_{S}$ determine the entirety of their respective spacetimes. In fact, the past domain of dependence for $\mathcal{R}_{QS}$ does not extend outside of $\mathcal{R}_{QS}$, and so the past is significantly underdetermined. 

Why can't we just select a region which does contain a Cauchy surface? Because this region would not be isometric, even approximately, to any region of collapse-Schwarzschild, and so we won't be able to use approximation to justify performing the derivation on collapse-Schwarzschild and transferring the result of the derivation back over to evaporation-Schwarzschild. Given this, let us see how each of the derivations fair.

Consider Hawking's derivation: it depends on global spacetime structure in the sense of an infinite past prior to collapse that is stationary, and an infinite future after collapse that is stationary, and a non-stationary intervening period. He writes: ``To understand how the particle creation can arise from mixing of positive and negative frequencies, it is essential to consider not only the quasi-stationary final state of the black hole but also the time-dependent formation phase." \cite[p. 204]{Hawking:1975vcx} $\mathcal{R}_{S}$ contains none, or at most very little, of this requisite structure. For example, any given $\mathcal{R}_{QS}$ need not intersect the collapse region; indeed, the majority not, as demonstrated in figure \ref{fig:Quasi-stationary}. Thus, the approximately isometric region $\mathcal{R}_{S}$ will also not intersect the non-stationary collapsing matter and so will have insufficient structure to carry out Hawking's derivation.

In order to recover the necessary structure, one needs to justify moving from  $\mathcal{R}_{S}$ to a spacetime with the global structure of collapse-Schwarzchild, perform the derivation on the global structure, and then make inferences from the global derivation back to the slice. Even if we assume that the first problem discussed above is solved and so such an inference is permissible, the inference still fails because, as per the second problem above, $\mathcal{R}_{S}$ does not contain a Cauchy surface for collapse-Schwarzschild. Therefore, even if one could justify using the very different global structure to draw inferences about $\mathcal{R}_{QS}$, not enough of the global structure is included in the MGHD of $\mathcal{R}_{S}$ to carry out Hawking's derivation. Therefore, Hawking's approximation regime fails for Hawking's derivation.\footnote{Given the equations governing quantum fields on curved spacetimes are local, states on local regions of spacetime should be insensitive to global structure. However, this reasonable argument faces the same challenges stated above. If we wish to justify the existence of Hawking radiation on a local region of the spacetime using Hawking's derivation, we must use the global structure Hawking uses, but this global structure seems to violate the very laws of quantum field theory on curved spacetime use just appealed to due to backreaction arguments. Making such an locality-based argument work would be a interesting and valuable contribution to our understanding of global idealizations. I am grateful to an anonymous referee for pushing me on this point.}

Turning to Fredenhagen and Haag's derivation: it was designed to not require the propagation of modes through the non-stationary collapse region, so the failure to recover this structure in $\mathcal{R}_{S}$ is not problematic. However, the asymptotic time limit is not recovered in $\mathcal{R}_{S}$; in fact the time over which the detector can be propagated is even shorter than in MGHD $\Ip$. Therefore, the modes cannot accumulate arbitrarily close to the horizon, as is needed in  Fredenhagen and Haag's derivation. However, perhaps the result is recovered approximately with this limited time evolution. Moreover, $\mathcal{R}_S$ does determine the entire future of the spacetime, so if we can overcome the first problem above and justify using the MGHD of $\mathcal{R}_{S}$, we can in fact recover the asymptotic time-limit. 

Hence, the approximation regime holds reasonable promise of succeeding for Fredenhagen and Haag's derivation. Nonetheless, it needs to be shown that either the use of the MGHD of $\mathcal{R}_S$ is justified, or $\mathcal{R}_S$ admits sufficiently long stationary worldlines to allow modes to accumulate sufficiently close to the horizon that the singularity structure of the quantum field will dominate. Neither of these are trivial, but neither seems implausible either. 

Finally, the algebraic approach: one begins by restricting the algebra of observables on collapse-Schwarzschild of mass $m$ to an algebra on $\mathcal{R}_{S}$. One then infers the vacuum state on this algebra by restricting the Unruh vacuum to $\mathcal{R}_{S}$. One then claims that the algebra and vacuum state on $\mathcal{R}_{QS}$ is approximately that of $\mathcal{R}_{S}$.

The central challenge for this approach is again the failure of $\mathcal{R}_{S}$ to be Cauchy. This means the uniqueness of the state on $\mathcal{R}_{S}$ will probably not hold. Without uniqueness, we can't guarantee the state on $\mathcal{R}_{S}$ is the restriction of the state on collapse-Schwarzschild. Moreover, the states on each $\mathcal{R}_{QS}$ must be smoothly joined together, and therefore one needs to understand how the approximation changes the state, if only slightly.\footnote{Work has begun to formulate algebraic QFT on non-globally hyperbolic spacetimes, for example \citeA{janssen2022quantum}.}

I do not claim that I have exhausted the possibilities and difficulties for Hawking's proposal. Nor do I claim that Hawking's proposal exhausts the possible approximation regimes. I simply claim that, as of yet, this approach hasn't been completely worked out for any of the derivations. Moreover, if the approximation regime is worked out for one derivation, say Fredenhagen and Haag's, then the idealization paradox remains for the others, and so interesting open questions remain. The goal of this section has been to emphasise that the inference from that fact that evaporation is slow to the claim that the derivations go through approximately unaffected is non-trivial. 

In the next section I consider whether we can weaken the premises of the derivations to deidealize them. 

\subsection{Essential Structure}\label{ssec:Implicit_structure}

The idealization paradox arises because some derivation uses a set of properties $X$ with which to derive Hawking radiation, and then one finds that evaporation spacetimes don't instantiate the set of properties $X$. However, suppose that one could show that the derivation in fact did not require the complete set of properties $X$ but only some subset of $X$, call it set $Y$, the essential structure. Suppose further that evaporation spacetimes could instantiate the essential structure $Y$. Then the inconsistency would be resolved. Moreover, the essential structure that goes into deriving Hawking radiation would have been identified, and the surplus structure is stripped away. 

The task of identifying this essential structure is undertaken in \citeA{VISSER_2003, barcelo2011hawking, barcelo2011minimal}. \citeA{VISSER_2003} argues that only three features are required for a derivation of Hawking radiation: an apparent horizon, non-zero surface gravity of the apparent horizon, and slow evolution. Therefore, using these as the set of properties $Y$ could potentially resolve the idealization paradox. Going further, \citeA{barcelo2011hawking, barcelo2011minimal} argue that Hawking-like radiation will occur whenever there is a continuous function mapping an affine parameter on future null infinity to that on past null infinity and the `adiabatic condition' is satisfied.\footnote{The authors argue that Hawking radiation will occur whenever the affine parameters, $U$ and $u$, of the null generators of $\Ip$ and $\If$ are approximately related by an exponential redshift such that $U \approx U_* - Ae^{-\kappa_* u}$ as $u \rightarrow \infty$. $U_*$ need not be the location of an event horizon but only a `best estimate' at $\If$. $\kappa_*$ is the value of the `peeling' function, $\kappa(u) = -\Ddot{p}(u)/\dot{p}(u)$ at $U_*$ where $p(u)$ is function $U = p(u)$, and $A$ is an arbitrary constant (see also \citeA{hu1996hawking}). The authors argue this condition will be satisfied whenever $|\dot{\kappa}(u_*)| \ll \kappa(u_*)^2$. This `adiabatic condition' is essentially a slow evolution condition. For details see \citeA{barcelo2011hawking}.} In \citeA{barcelo2011hawking} the authors show how these conditions, with added assumptions about the QFT, can be used to derive the Bogoliubov coefficients, making explicit the relationship between their minimal conditions and Hawking's derivation of Hawking radiation. 

Deidealization via essential structure derivations is strikingly different to that via the approximation regime. Whereas Hawking sought to find stationary structure within a non-stationary spacetime, these derivations do away with the need for quasi-stationary regions, and instead provide a derivation which would be successful on the global structure of an evaporation spacetime. Both \citeA{VISSER_2003} and \shortciteA{barcelo2011hawking, barcelo2011minimal} require the black hole to evolve slowly, but they do not use this slow evolution to approximate stationarity. By using a different deidealization method, different lessons are drawn. For example, given the very minimal structure used in these derivations, it can be argued that they point towards a kinematic interpretation of Hawking radiation, \textit{contra} the dynamical picture given in \citeA{Hawking:1975vcx}. Moreover, these derivations don't require an event horizon or Killing horizon to form. Similar lessons to those from an approximation regime can be learned here also, for example the spectrum derived is only approximately thermal, and the spectrum can be derived away from the asymptotic future (that is, before the retarded time coordinate goes to infinity).

I do not claim that these derivations face no difficulties, but only that they are very promising candidates for resolving the idealization paradox. A full analysis will be carried out in the future of the project. There is also a semantic issue of what one takes to be the referent of `Hawking radiation' which I ignore here, emphasising only that resolutions to the paradox modify: i) what one takes to be required for something like Hawking radiation to occur, and ii) what is observed at $\If$. On a cautious note, it is not clear that one can distinguish between radiation due to the Unruh effect and radiation due to the Hawking effect with these derivations. Although the Unruh effect and Hawking radiation are closely related phenomena, they are not the same \cite{Earman2011-EARTUE}. If one cannot distinguish between the two a derivation may have insufficient structure. However, this does not seem to me a serious obstacle to these derivations resolving the paradox, but rather an obstacle to the full interpretation of the Hawking effect.

The papers discussed here are not the only candidates for essential structure resolutions. Quantum tunnelling approaches, for example \citeA{Parikh_2000}, give a local dynamical account of Hawking radiation. A resolution to the paradox by these derivations would tell a different story. Firstly, they would retain a dynamical ontology for Hawking radiation. Secondly, they would point to lessons about the encoding of local structure by global structure in semi-classical gravity. Unpacking this second point, the definition of a black hole is global and Hawking's derivation is global, but if a resolution of the paradox along the lines of a local dynamical account is the correct one, we might learn that this global structure is a red herring, and it just encodes local structure that in ways that are, at times, opaque. 

The goal of this section has been three-fold: 1) To highlight the differences between different deidealization strategies, 2) To emphasise there is an alternative to approximation regimes which make an inference from slow-evaporation to unaffected derivations, 3) To highlight the importance of research programmes such as that undertaken by \citeA{VISSER_2003, barcelo2011hawking, barcelo2011minimal}. I do not claim that the paradox is definitely solved, or even necessarily solved by an essential structure deidealization, but rather that this is a promising option with many lessons to be learnt.

Summarising, deidealization can follow multiple different routes and these routes have varying degrees of success. Indeed, the success of a particular deidealization need not be homogeneous across derivations. I only take the quantum gravity route to be completely impotent. Hawking's approximation regime fails for Hawking's derivation, but prospects for success are better for Fredenhagen and Haag's derivation, and other approximation regimes may fare better. The essential structure research programme is very promising, in particular for deidealizing Hawking's derivation. The lessons we draw from these varying approaches to deidealization depend upon the type of deidealization, and the details of how the deidealization operates.

\section{Conclusion}

\epigraph{Paradoxes are just the scar tissue. Time and space heal themselves up around them and people simply remember a version of events which makes as much sense as they require it to make.}{Douglas Adams, \textit{Dirk Gently's Holistic Detective Agency}}

I have argued that Hawking's derivation of Hawking radiation, along with Fredenhagen and Haag's and the algebraic approach, are all evaporation-inconsistent. They are carried out on collapse-Schwarzschild but cannot be carried out on evaporation-Schwarzschild. By throwing away the spacetime used to derive the phenomenon, we throw away the very ladder we are standing on, and come tumbling back to inconsistency. There are reasonable (and some unreasonable) paths towards deidealizing the derivations involved, and thus reason to believe that the paradox is just scar tissue from the messy process of scientific development. Presumably, there is a resolution along the lines of an approximation regime or essential structure derivation which will teach us why the inconsistent derivations worked so well, and what they really represent. If so, this will be another victory for the dispensabilists in the idealization literature, and a particularly striking one given that the idealizations used in Hawking radiation derivations were not simply false, but inconsistent. The differences between the different possible deidealizations emphasises the non-triviality of the deidealization project, and the variety of lessons that may be learnt. Most excitingly, deidealizing these derivations may remove the chaff from the conceptual framework of Hawking radiation and give a clear ontological picture of Hawking's eponymous discovery. Such lessons are the fruits of paying close attention to, and resolving, the idealization paradox; fruits won as reward for not settling with the unjustified inference from slow-evaporation to unaffected derivations. No matter what road we take, we are bound to learn something interesting. 

\section*{Acknowledgements}

I am very grateful to Bryan Roberts for his fantastic guidance and support in the development and refinement of this paper. I would also like to thank the participants in the Philosophy of Physics Bootcamp and in particular Erik Curiel for careful and helpful reading of an earlier draft, as well as simulating discussion on related topics. Finally, I am indebted to Bruno Arderucio, Jeremy Butterfield, Saakshi Dulani, Sam Fletcher, Henrique Gomes, Sean Gryb, Nick Huggett, Klaas Landsman and Karim Th\'ebault, and to three anonymous referees who helped improve this paper greatly. Finally my thanks to audiences at: the University of Bristol Philosophy of Physics seminar, the Cosmology and Quantum Gravity Beyond Spacetime conference at Western University, the Golden Wedding of Black Holes and Thermodynamics conference, the Harvard Black Hole Initiative Foundations seminar and the annual conference of the Philosophy of Physics Group of the German Physical Society in Berlin.

\bibliographystyle{apacite}
\bibliography{bibliography}

\end{document}